\documentclass[
aps
 ,twocolumn, nofootinbib
]{revtex4}
\usepackage{tikz}
\usepackage{adjustbox}
\usetikzlibrary{arrows, hobby}
\usepackage{wrapfig}
\usepackage{graphicx}
\usepackage{amsmath}
\usepackage{amssymb}
\usepackage{hyperref}
\usepackage{ mathrsfs }
\usepackage{indentfirst}
\usepackage{braket}
\usepackage{array}
\usepackage{color}
\usepackage{empheq}
\usepackage{lipsum,adjustbox}
\usepackage[outdir=./]{epstopdf}
\usepackage{booktabs}
\usepackage{tabularx}
\usetikzlibrary{calc,positioning}
\tikzstyle{beamsplitter}=[fill=blue, fill opacity=0.2]

\begin{document}

\title{Feasibility
 Assessment For Practical Continuous Variable Quantum Key Distribution Over The Satellite-to-Earth Channel}

\author{S. P. Kish$^1$, E. Villase\~{n}or$^1$, R. Malaney$^1$, K. A. Mudge$^2$ and K. J. Grant$^2$}\affiliation{$^1$School of Electrical Engineering and Telecommunication, \\
	University of New South Wales, Sydney, New South Wales 2052, Australia. \\
	$^2$Defence Science and Technology Group, Edinburgh, SA 5111, Australia.}

\date{\today}

\begin{abstract}
Currently, quantum key distribution (QKD) using continuous variable (CV) technology has only been demonstrated over short-range terrestrial links. Here we attempt to answer whether CV-QKD over the much longer satellite-to-Earth channel is feasible. To this end, we first review the concepts and technologies that will enable CV-QKD over the satellite-to-Earth channels. We then consider, in the infinite key limit, the simplest-to-deploy QKD protocols, the coherent state (CS) QKD protocol with homodyne detection and the CS-QKD protocol with heterodyne detection. We then focus on the CS-QKD protocol with heterodyne detection in the pragmatic setting of finite keys, where complete security against general attacks is known. We pay particular attention to the relevant noise terms in the satellite-to-Earth channel and their impact on the secret key rates. In system set-ups where diffraction dominates losses, we find that the main components of the total excess noise are the intensity fluctuations due to scintillation, and the time-of-arrival fluctuations between signal and local oscillator. We conclude that for a wide range of pragmatic system models, CS-QKD with information-theoretic security in the satellite-to-Earth channel is feasible.
\end{abstract}
\maketitle

%

%


\
%
\section{Introduction}
Quantum key distribution (QKD) provides information-theoretic secure key distribution between two parties. Local data sent by a sender, Alice, is encrypted using a key that is guaranteed by quantum mechanics to be securely shared only with the receiving party, Bob. QKD is mainly implemented using optical technology, making it ideal for secure key distribution over a high bandwidth free-space optical (FSO) link.

In discrete variable (DV) QKD applied over fibre the quantum information is usually encoded in a DV of a single photon (e.g. time-bin qubit), with a secure range of order 400~km \cite{letter}. However, recently, DV-based QKD using polarization as the DV was demonstrated in a free-space satellite-to-ground channel up to a distance of $1200$~$\text{km}$ \cite{liao}. In \cite{liao}, a quantum link was established from the low-Earth-orbit (LEO) {\it Micius} satellite to the Xinglong ground station. This exciting development is a major step towards realizing global scale secure quantum communications using low-orbit satellites. Indeed, quantum communication
through satellite channels is anticipated to be one of the core technologies enabling the so-called quantum internet \cite{bed}.

However, it is uncertain whether DV or continuous variable (CV) multiphoton technologies (or a combination of CV-DV) will prevail in free-space optical quantum communication. CV protocols have the advantage of high-rate, efficient and cost-effective detection (using homodyne and heterodyne detectors) in comparison to the sophisticated and expensive single-photon detectors used for DV protocols. CV-QKD is also perhaps more compatible with current classical wireless communications technologies. Initially, CV-QKD was proposed with discrete and Gaussian encoding of {\it squeezed} states \cite{ralph, hill, reid, cerf}. Thereafter, Gaussian-modulated CV-QKD with {\it coherent} states was soon developed \cite{gmod, gross, gross2}.

The simplest-to-deploy  CV-QKD protocols are the GG02 protocol introduced in 2002 by Grosshans and Grangier using a homodyne detector \cite{gross, gross2}, and its heterodyne variant (``the no-switching" protocol) \cite{weedbrook}. These both involve preparation of a coherent state (CS) using Gaussian modulation \cite{cerf, gmod}. More specifically, the quadratures $X$ and $P$ of the coherent state are randomly modulated according to a Gaussian distribution \cite{gmod}. In this paper, we refer to GG02 and the no-switching protocol as the CS-Hom protocol and CS-Het protocol, respectively. We state them as the CS-QKD protocols when we refer to them collectively. The CS-QKD protocols have been successfully deployed in fibre with non-zero quantum keys distributed over distances of order $100~\text{ km}$ \cite{gmod, xwang, jou}. 

The question we address in this work is whether the simplest-to-deploy CV-QKD protocols, namely the CS-QKD protocols, will be viable in a real-world LEO satellite-to-Earth channel. Although extensive theoretical work on CV-QKD through terrestrial FSO channels has been done in recent years, e.g. \cite{ber, usenko, heim, qu, gong,papan, rupert, derk}, the only real-world FSO deployment of any CV-QKD protocol  has been over the short distance of $460~\text{ m}$ \cite{polarize}.
The larger losses due to diffraction over the much longer satellite-to-Earth channel (as well as disparate turbulence effects) \cite{neda1}, render any extrapolation
of the results in \cite{polarize} far from obvious. This is compounded by the fact that \cite{polarize} deployed a different form of CV-QKD, namely, a CS unidimensional CV-QKD protocol.



Our contributions in this work can be summarized as follows.
We identify the most important factors contributing to the total excess noise of the CS-QKD protocols in the satellite-to-Earth channel. We pay particular attention to the noise contributions arising from scintillation and contributions arising from time-of-arrival fluctuations between the transmitted local oscillator (LO) and the quantum signal. Using our determination of the total excess noise we then consider a lower bound on the secure key rates anticipated for the CS-QKD protocols deployed over the satellite-to-Earth channel. We will consider secure rates in the asymptotic signalling limit for both of the CS-QKD protocols, and in the finite key limit for the CS-Het protocol.\footnote{Currently, a formal proof of security in the finite limit for the CS-Hom protocol is unavailable.}

%

The rest of this work is as follows.  In Section \ref{security} we review the security models required for calculating the secret key rate of the CS-QKD protocols in a lossy channel.  In Section \ref{coherent} we introduce the system model of the CS-QKD protocols in the satellite-to-Earth channel, and review well-known contributions to the total excess noise from the channel and detectors. In Section \ref{simulate}, we simulate the relative intensity fluctuations and time-of-arrival fluctuations. In Section \ref{impact}, we investigate the impact on the secret key rate of each noise term. In Section \ref{qhacking}, we discuss the security of the CS-QKD protocols in the satellite-to-Earth  channel and discuss alternative protocols. In Section \ref{discuss}, we discuss our work in the context of other CV-QKD protocols. Finally, we conclude and summarize our findings in Section \ref{conclusion}.



\section{Security analysis of CS-QKD in a lossy channel}
\label{security}

In our security model of the CS-QKD protocols in the satellite-to-Earth channel, we assume the eavesdropper Eve has full access to the quantum channel and performs the most general (coherent) attack. The general attack is when Eve prepares an optimal global ancilla state, which may not necessarily be separable. Eve can prepare any ancillary states to interact with the signal states and make measurements. In particular, Eve can have access to a quantum memory, which can store the signal states until she learns information during the classical post-processing. In our security model, we perform the analysis in the entanglement based (EB) version of the CS-QKD protocols, even though in deployment we will assume the equivalent prepare and measure (PM) schemes \cite{laud,virtual}. In deployment, it is straightforward to convert measurements from one scheme to the other \cite{laud}.

Information-theoretic security against the general attack for the CS-QKD protocols was proven in the {\it asymptotic limit} where an infinite key was assumed \cite{diam}. The general attack was proven to reduce to the collective attack using the de Finetti representation theorem for infinite dimensions \cite{renner}. However, in practice, we need to consider the secret key rate in the finite limit which can only be $\epsilon$-secure, where $\epsilon$ is the probability of failure. Information-theoretic security of the CS-Hom protocol in the finite regime is still an open question.
However, by exploiting a new Gaussian de Finetti reduction method, a composable security proof against general attacks for the {\it CS-Het} protocol was recently proven \cite{lev2}. 

\subsection{Asymptotic limit}
In this section, we review the well-known secret key rates under general attacks in the asymptotic limit (which always provides key rates higher than those obtained under general attacks in the finite limit). We will follow the formalism provided in \cite{foss} and references therein.
The covariance matrix describing the Gaussian modulated coherent state sent from Alice (A) to Bob (B) has the form,
\begin{equation}
\gamma_{\rm AB}=
\begin{pmatrix}
a\mathbb{I} & c\sigma_{z} \\ c\sigma_{z} & b\mathbb{I}
\end{pmatrix},
\label{covariance}
\end{equation}
where $\mathbb{I}=\text{diag}(1,1)$ is the unity matrix and $\sigma_{z}~=~\text{diag}(1,-1)$ is the Pauli matrix. For the Gaussian modulated CS-QKD protocols, the coefficients of the covariance matrix are
\begin{equation}
\begin{split}
a&=V_A+1, \\
b&=V_B=\frac{\eta_d T}{\mu} (V_A+\chi(T,\eta_d)+1), \\
c&=\sqrt{\frac{\eta_d T}{\mu}} \sqrt{V_A^{2}+2V_A},
\end{split}
\label{covariancematrix}
\end{equation} where $V_A$ is Alice's modulation variance of the quadrature operators, $V_B$ is Bob's measured quadrature variance, $0 \le T \le 1$ is the transmissivity of the channel, and $\eta_d$ is the detector quantum efficiency.\footnote{We implicitly assume in all calculations that the vacuum noise is normalized to $1$ by setting $\hbar=2$.} The CS-Hom and CS-Het protocols correspond to $\mu=1$ and $\mu=2$, respectively. The term $\chi(T, \eta_d)$ is defined as all noise terms other than vacuum noise, expressed in vacuum units. 
In the literature it is common to distinguish between noise, $\chi_d(\eta_d)$, arising from the detector and noise, $\chi_{ch}(T)$, arising from the channel transmission where \cite{foss},
\begin{equation}
\chi(T,\eta_d)=\chi_{ch}(T)+\frac{\chi_d (\eta_d)}{T},
\label{totalexcess}
\end{equation} and where $\chi_{ch}(T)$ is given by
\begin{equation}
\chi_{ch}(T)=\frac{1-T}{T}+\xi_{ch}.
\label{chT}
\end{equation}
Here, $\xi_{ch}$ is the channel excess noise from various sources and $(1-T)/T$ is due to channel loss. $\chi_{d}(\eta_d)$ is given by
\begin{equation}
\chi_d(\eta_d)=\frac{\mu-\eta_d}{\eta_d}+\frac{\mu \xi_{d}}{\eta_d},
\label{dT}
\end{equation} where $\xi_{d}$ is the detector excess noise and $(\mu-\eta_d)/\eta_d$ is noise due to detector losses.

The secret key rate\footnote{When we refer to ``secret key rate" in this paper, we actually mean a lower bound on the rate.} against general attacks (in bits/pulse) under reverse reconciliation in the asymptotic limit (infinite key length) is given by \cite{dev, scar, foss}
\begin{equation}
K=\beta I_{AB}-S_{BE},
\label{keyrate1}
\end{equation} where $I_{AB}$ is the mutual information between Alice and Bob, $0 \le \beta \le 1$ is the reconciliation efficiency, and $S_{BE}$ is the upper bound to the Holevo information between Eve and Bob. In this work we will only consider reverse reconciliation where Bob sends correction information to Alice who corrects the bit values in the key (derived from her quadrature measurement) \cite{gross1}. The mutual information $I_{AB}$ for the CS-QKD protocol is given by
\begin{equation}
I_{AB}=\frac{\mu}{2} \log_2{\frac{V_A+1+\chi(T,\eta_d)}{1+\chi(T,\eta_d)}}.
\label{mut}
\end{equation}
Note that from this point forward, we simplify the nomenclature with $\chi(T,\eta_d):=\chi$, $\chi_{ch}(T):=\chi_{ch}$ and $\chi_d(\eta_d):=\chi_d$. Note also, the~mutual~information~becomes 
\begin{equation}
I_{AB}~=~\frac{\mu}{2} \log_2{\frac{\frac{\eta_d T}{\mu}(V_A+\xi_{ch})+\xi_{d}+1}{\frac{\eta_d T}{\mu}(\xi_{ch})+\xi_{d}+1}},
\end{equation} and in the limit $T\rightarrow 0$,~$I_{AB(T\rightarrow 0)}=\frac{\mu}{2}\log_2{\frac{1+ \xi_{d}}{1+ \xi_{d}}}=0$, as expected.

We use the trusted model throughout the paper, where Eve can only have access to the channel excess noise $\chi_{ch}$. Consequently, in the Holevo information shared between Eve and Bob in (\ref{hol}), we only consider $\chi_{ch}$.
Eve's information after Bob's measurement is the upper bound on the Holevo information,
\begin{equation}
S_{BE}=S(E)-S(E|B),
\label{hol}
\end{equation}
where $S(E)$ is the von Neumann entropy of Eve's state before the measurement on mode $B$ (Bob's mode) and $S(E|B)$ is conditioned on Bob's measurement outcome. Eve is assumed to purify the transmitted state $AB$ resulting in a pure Gaussian state. It was shown that for a given covariance matrix describing a Gaussian state, Gaussian attacks are the most optimal attacks that minimize the key rates \cite{garcia}. Therefore, Eve's entropy is
\begin{equation}
S(AB)=G(\frac{\lambda_1-1}{2})+G(\frac{\lambda_2-1}{2}),
\label{AB}
\end{equation}
where $G(x)=(x+1) \log_2{(x+1)}-x\log_2{x}$ and $\lambda_{1,2}$ are the symplectic eigenvalues of the covariance matrix $\gamma_{AB}$. For CS-QKD protocols, these are given by \cite{foss}
\begin{equation}
\begin{split}
\lambda_{1,2}&=\sqrt{\frac{A' \pm \sqrt{A'^2-4 B'}}{2}},  \text{where} \\
A'&=(V_A+1)^2 (1-2 T)+2T+T^2(V_A+1+\chi_{ch}), \\
B'&=T^2((V_A+1) \chi_{ch}+1)^2,
\end{split}
\label{lambda1}
\end{equation} and the entropy $S(E|B)$ is
\begin{equation}
S(E|B)=G(\frac{\lambda_3-1}{2})+G(\frac{\lambda_4-1}{2})+G(\frac{\lambda_5-1}{2}),
\label{EB}
\end{equation} where $\lambda_{3,4,5}$ are the symplectic eigenvalues of the conditional covariance matrix characterizing the state after Bob's measurement. The eigenvalues $\lambda_{3,4}$ are given by,
\begin{equation}
\lambda_{3,4}=\sqrt{\frac{C' \pm \sqrt{C'^2-4 D'}}{2}},
\label{lambda2}
\end{equation} where for the CS-Hom protocol $C'$ and $D'$ are given by
\begin{equation}
\begin{split}
C'_{hom}&=\frac{A' \chi_d+(V_A+1) \sqrt{B'}+T(V_A+1+\chi_{ch}) }{T(V_A+1+\xi)}, \\
D'_{hom}&=\sqrt{B'} \frac{V_A+1+\sqrt{B'} \chi_d}{T(V_A+1+\chi)},
\end{split}
\label{lambda3}
\end{equation} respectively, and for the CS-Het protocol $C'$ and $D'$ are given by
\begin{equation}
\begin{split}
C'_{het}&=\frac{1}{T^2(V_A+1+\chi)} [A' \chi_d^2+B'+1 \\
&+2 \chi_d ((V_A+1)\sqrt{B'}+T(V_A+1+\chi_{ch}))\\
&+2 T (V_A^2+2V_A)], \\
D'_{het}&=\bigg(\frac{V_A+1+\sqrt{B'}\chi_d}{T(V_A+1+\chi)}\bigg)^2,
\end{split}
\label{lambda4}
\end{equation} respectively. Note, $\lambda_5=1$ for both protocols. Using (\ref{keyrate1}) and (\ref{AB})-(\ref{lambda4}), it is then straightforward to calculate the secret key rate in the asymptotic limit.

\subsection{Finite key size effects}
We now consider the practical security of the CS-QKD protocols using finite-size analysis. We will assume the small finite processing errors associated with the Gaussian modulation of the coherent states can be ignored. We do note, however, that an information-theoretic validation of this assumption does not yet exist \cite{takaya}. In the security analysis, we consider the Gaussian collective attacks with $\epsilon$-security that was first introduced in \cite{lev, imperfect} and extended to general attacks in \cite{lev2, levc, lupo, neda2020}. In the infinite limit, the knowledge of the relevant parameters are exact, but in the finite limit, the parameters are estimated with a finite precision. This precision is related to the probability $\epsilon_{PE}$ the true values are not inside the confidence interval calculated from the parameter estimation procedure \cite{lev, imperfect}. In the finite limit, the secret key rate
in bits/pulse is given by \cite{levc, lupo}
\begin{equation}
K= \frac{n}{N} [\beta I_{AB}-S^{\epsilon_{PE}}_{BE}]-\frac{\sqrt{n}}{N}\Delta_{AEP} (n)-\frac{2}{N} \log_2{\frac{1}{2 \epsilon}},
\label{finitekey}
\end{equation}
where $\epsilon$ is the total failure probability of the protocol, $S^{\epsilon_{PE}}_{BE}$ is the upper bound of the Holevo information taking into consideration the finite precision of the parameter estimation, $N$ is the total number of symbols sent, and $n=N-n_e$, where $n_e$ is the number of symbols used for parameter estimation. $\Delta_{AEP}(n)$ is given by \cite{levc, lupo}
\begin{equation}
\begin{split}
\Delta_{AEP} (n)&=(d+1)^2+4(d+1) \sqrt{\log_2(2/\epsilon_s)}\\
&+2\log_2(2/(\epsilon^2 \epsilon_s))+4 \epsilon_s d/(\epsilon \sqrt{n}),
\end{split}
\label{finitelog}
\end{equation} where $d$ is the discretization parameter\footnote{$d$ is the bits of precision encoded by the symbol. In this work, we set $d=5$ as in \cite{lupo}. } and $\epsilon_s$ is a smoothing parameter corresponding to the speed of convergence of the smooth min-entropy. In the finite-size regime, one is limited to $\epsilon$-security where $\epsilon=\epsilon_{EC}+2\epsilon_s+\epsilon_{PA}+\epsilon_{PE}$, where $\epsilon_{PA}$ is the failure probability of the privacy amplification procedure, and $\epsilon_{EC}$ is the failure probability of the error correction. The parameters $\epsilon_s$ and $\epsilon_{PA}$ can be optimized computationally \cite{lev}.

To account for the finite statistics and obtain the value of $S^{\epsilon_{PE}}_{BE}$, the previous covariance matrix
(\ref{covariancematrix}) for Alice and Bob becomes \cite{lev, imperfect},
\begin{equation}
\label{eq4}
\begin{split}
&\gamma^{\epsilon_{PE}}_{\rm AB}\\
&= \begin{pmatrix}
(V_A+1) \mathbb{I} & \sqrt{\frac{T_{min}}{2} }\sqrt{V_A^{2}+2V_A}\sigma_{z} \\ \sqrt{\frac{T_{min}}{2}}\sqrt{V_A^{2}+2V_A}\sigma_{z}
& (\frac{T_{min}}{2}(V_A+\xi_{max})+1+\xi_{d}))\mathbb{I}
\end{pmatrix},
\end{split}
\end{equation}
where $T_{min}$ and $\xi_{max}$ are the minimum and maximum values of $T$ and $\xi_{ch}$, respectively.
\begin{figure*}[t!]
\centering
\includegraphics[scale=0.5]{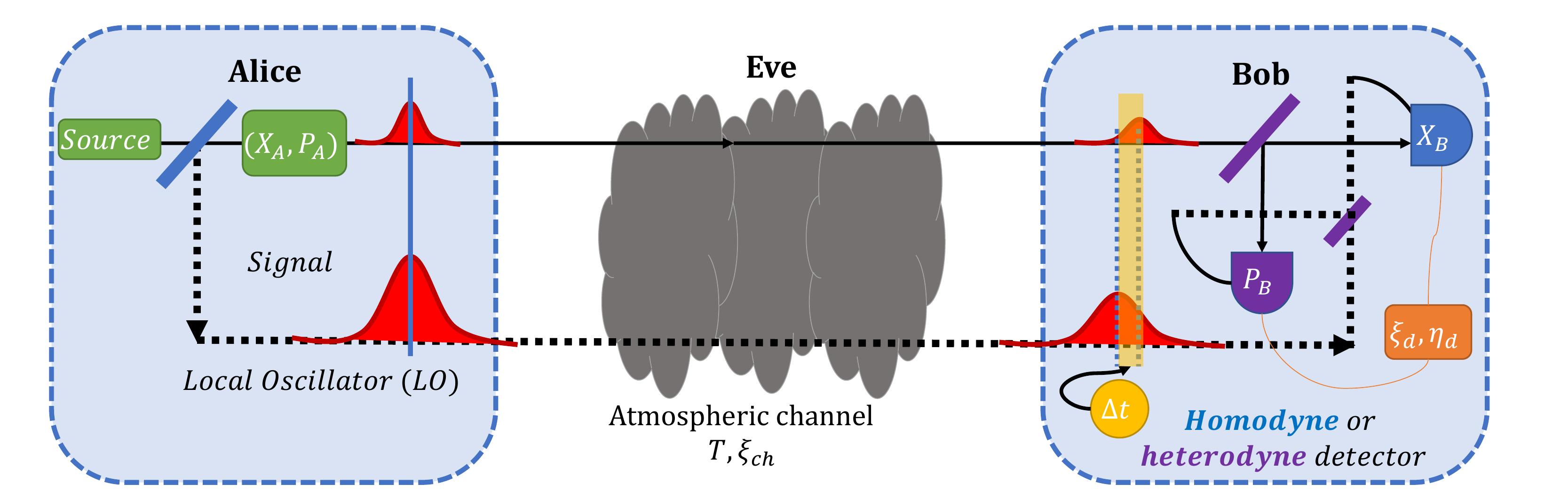}
\caption{The PM CS-QKD protocols in the satellite-to-Earth channel. A local oscillator (LO) is sent with the signal by Alice which passes through a lossy channel of transmissivity $T$ and channel excess noise $\xi_{ch}$. Bob receives the signal and the LO which he uses to perform the homodyne or heterodyne detection. The former is represented by a $X_B$ quadrature measurement (blue), and the latter by two balanced beamsplitters (purple) and two quadrature measurements (blue $X_B$ and purple $P_B$). The heterodyne/homodyne detector efficiency is $\eta_d$ and the detector excess noise $\xi_{d}$. An entirely equivalent EB scheme is available.}
\label{skrlossblock}
\end{figure*}

The confidence intervals for  $T$ and $\xi_{ch}$ can be calculated from known distributions of the estimators\footnote{In a real-world deployment the maximum-likelihood estimators can be computed from the measurement data to determine $T_{min}$ and $\xi_{max}$ \cite{lev, imperfect}.} $\hat{t}$ and $\hat{\sigma}^2$ to obtain the lower value of the $T$ interval given by \cite{lev, imperfect},
\begin{equation}
T_{min}=\bigg( \hat{t}-z_{\epsilon_{PE}/2} \sqrt{\frac{\hat{\sigma}^2}{n_e V_A}} \bigg)^2,
\label{tmin}
\end{equation} and the higher value of the $\xi_{ch}$ interval given by,
\begin{equation}
\xi_{max}=\bigg( \hat{\sigma}^2+z_{\epsilon_{PE}/2} \frac{\hat{\sigma}^2 \sqrt{2}}{\sqrt{n_e}}-1-\xi_{d} \bigg)/\hat{t}^2,
\label{emax}
\end{equation} where $z_{\epsilon_{PE}/2}$ satisfies $1-\text{erf}(z_{\epsilon_{PE}/2}/ \sqrt{2})=\epsilon_{PE}$ and $\text{erf}(x)$ is the error function defined as $\text{erf}(x)~=~\frac{2}{\sqrt{\pi}} \int^x_0 e^{-t'^2} dt'$. From a theoretical perspective and what we do in this work, we can set the expectation values of the estimators to
\begin{equation}
\begin{split}
E[\hat{t}]&=\sqrt{\eta_d T}, \\
E[\hat{\sigma}^2]&=T\eta_d \xi_{ch}+1+\xi_{d}.
\end{split}
\end{equation}
Using these values, we can compute $T_{min}$ and $\xi_{max}$.
To determine $S_{BE}^{\epsilon_{PE}}$ in (\ref{finitekey}) for the CS-Het protocol, we use these values in equations (\ref{hol})-(\ref{lambda2}) and (\ref{lambda4})(i.e. setting $T=T_{min}$ and $\xi_{ch}=\xi_{max}$ in equations (\ref{totalexcess})-(\ref{dT})). The mutual information $I_{AB}$ in (\ref{finitekey}) is calculated as done before in (\ref{mut}) using $T$ and $\xi_{ch}$. Putting all this together, to calculate the secret key rate with finite size effects in (\ref{finitekey}), we make use of the aforementioned quantities $S_{BE}^{\epsilon_{PE}}$, $I_{AB}$, and $\Delta_{AEP}(n)$ in (\ref{finitelog}).

In going from optimal Gaussian collective attacks to general attacks, only the CS-Het protocol is $\epsilon'$-secure against general attacks with $\epsilon'=\frac{\kappa^4}{50} \epsilon$, where \cite{lev2}
\begin{equation}
\begin{split}
\kappa&=\text{max}\{1,n(d_A+d_B)(1+2\sqrt{[(\ln(8/\epsilon))/2n]})\\
&+(\ln(8/\epsilon)/n)(1-2\sqrt{[\ln(8/\epsilon)/2k]})\},
\end{split}
\end{equation} where $d_A$ and $d_B$ are the average photon numbers of Alice and Bob's modes respectively, and $k$ is the number of modes \cite{lev2}. However, $\kappa\approx n$ and thus $\epsilon \approx \frac{50}{n^4} \epsilon'$. We choose $\epsilon'=10^{-9}$ and number of symbols used for parameter estimation\footnote{In the rest of the paper, unless stated otherwise, we choose $n_e=n=10^{12}$, implying $n/N=0.5$.} $n_e=n=10^{12}$ which can be obtained in minutes for a source pulse rate of $100$~MHz. To obtain the secret key rate under general attacks, it is enough to analyze the security against Gaussian collective attacks with
$\epsilon=10^{-55}$ and $\epsilon_{EC}=\epsilon_s=\epsilon_{PA}=\epsilon_{PE}$ \cite{lev2}.
\section{Satellite-to-Earth CS-QKD}
\label{coherent}



We present our system model in Fig.~\ref{skrlossblock}. Essentially, a Gaussian modulated coherent state is prepared on the satellite (Alice) and measured at the ground station (Bob) using homodyne or heterodyne detection. At Alice's location in a LEO satellite at altitude $H$, a strong laser source generates a coherent state which is divided by an asymmetrical beamsplitter into the local oscillator (LO) and the signal path. The laser source generates pulses at central frequency $\omega_0$ of width $\tau_0$ and a repetition rate $f_{rep}$. The laser beam is collimated by a transmitter aperture of diameter $D_T$. The amplitude and phase corresponding to $X$ and $P$ quadratures in the signal path are Gaussian modulated with the Gaussian distribution centred at $\braket{X}=\braket{P}=0$ with variance $V_A$.

The LO is then multiplexed with the signal (in a different polarization mode) and sent to Bob. Both the LO and signal are received by an aperture of diameter $D_R$. Using the LO, Bob monitors the transmissivity $T$ which has a probability density function PDF $P_{AB}(T)$ of the channel. The PDF is integrated up to a maximum possible value of $T=T_{max}$. In general, the form of this PDF is a function of many atmospheric parameters,  the transceiver apertures, and the distance between the transceivers. In the uplink (Earth-to-satellite) deep fades in the transmissivity can be anticipated, largely due to beam wander and beam deformation \cite{yu}. In this work, we adopt settings where the losses in the satellite-to-Earth channel are dominated by diffraction effects alone. That is, we assume the transmissivity is a constant. As we discuss later, reasonable receiver/transmitter apertures render such an assumption reasonable in the downlink channel.\footnote{Even in cases where non-constant transmissivity is present, channel post-selection using the LO can render a subset of the channels to lie within an effective constant transmissivity window. QKD rates can then be approximated via summation of rates arising from these subsets.} However, it is worth noting here that additional support for a constant transmissivity in the downlink comes from the phase-screen simulations of \cite{Eduardo}, which show highly-peaked transmissivity PDFs.\footnote{Note the QKD key rates of \cite{Eduardo} are based on the use of an LO generated directly at the receiver (not one transmitted with the signal), and therefore cannot be directly compared with the key rates reported here.}


In the CS-Hom protocol, Bob randomly chooses to either measure the quadratures $X$ or $P$ with the homodyne detector and later announces to Alice which quadrature he measured. In the CS-Het protocol, Bob measures both $X$ and $P$ using two detectors at the output of a balanced beamsplitter.
\label{noise}
\nolinebreak
\subsection{Noise Components}
\label{excess}
As shown in (\ref{totalexcess}), we separated the noise due to the channel $\chi_{ch}$ and the detector $\chi_{d}$. Bob's variance\footnote{In the limit of $T\rightarrow 0$, Bob's variance  of the quadrature operator reduces to $1+\xi_{d}$ as expected.} $V_B$ of the quadrature operator in (\ref{covariancematrix}) using the definitions in (\ref{chT}) and (\ref{dT}) is \begin{equation}
\begin{split}
V_B=&\frac{\eta_d T}{\mu}(V_A+\chi_{ch}+\frac{\chi_{d}}{T}+1) \\
&=\frac{\eta_d T}{\mu}(V_A+\xi_{ch}+\frac{\mu \xi_{d}}{\eta_d T})+1 \\
&=\frac{\eta_d T}{\mu}(V_A+\xi)+1,
\end{split}
\end{equation} where we define the total excess noise as \begin{equation}
\xi:=\xi_{ch}+\frac{\mu \xi_{d}}{\eta_d T}.
\label{ex}
\end{equation}
The channel excess noise\footnote{The channel excess noise is defined with respect to the input, and is thus multiplied by the transmissivity $T$ and $\eta_d$ in the covariance matrix.} is \cite{laud, filip, swang2}
\begin{equation}
\begin{split}
\xi_{ch}&=\xi_{ta}+\xi_{RIN, Atmos} +\xi_{Background}+\xi_{mod}\\
&+\xi_{RIN, LO}+\xi_{RIN, Signal},
\end{split}
\end{equation} where the contributions above are the time-of-arrival fluctuations $\xi_{ta}$, relative intensity noise (RIN) due to the atmosphere $\xi_{RIN, Atmos}$, background noise $\xi_{Background}$, modulation noise $\xi_{mod}$, RIN of the LO $\xi_{RIN, LO}$, and RIN of the signal $\xi_{RIN, Signal}$. We refer the reader to Table~\ref{table1} for a description of these noise components. We note that the time-of-arrival fluctuations $\xi_{ta}$ and RIN due to the atmosphere, $\xi_{RIN, Atmos}$, have not been determined in the satellite-to-Earth channel. An analysis of these two excess noise components is discussed in the next section.

Similarly, the detector excess noise\footnote{The detector excess noise is defined with respect to the output and hence $T$ and $\eta_d$ cancel out in the covariance matrix.} is \cite{ren, tao, laud, huangsci, chi, bqi}
\begin{equation}
\begin{split}
\xi_{d}&=v_{el}+\xi_{ADC}+\xi_{overlap}+\xi_{LO}+\xi_{Leak}
\end{split}
\end{equation} where the noise contributions listed are the electronic noise $v_{el}$, analogue-digital converter noise $\xi_{ADC}$, detector overlap $\xi_{overlap}$, LO subtraction noise $\xi_{LO}$, and LO-to-signal leakage $\xi_{Leak}$. We refer the reader to Table \ref{table1} for a fuller description of these noise components.

\begin{table*}[t!]
\centering
\begin{tabular}{|p{1.6cm} | p{5cm} | p{10.5cm} |}
 \hline
 $\xi_{ch}$ & Channel excess noise component & Description \\
 \hline\hline

$\xi_{ta}$ & {\it Time-of-arrival fluctuations} & $\xi_{ta}$ is the noise component due to the differential modifications between signal and LO pulses in the satellite-to-Earth channel. Predictions for this noise component in a terrestrial free-space channel were made in \cite{swang}, but not for the satellite-to-Earth channel. We will quantify this noise component in the next section. \\
\hline
$\xi_{RIN, Atmos}$& {\it RIN of LO due to atmosphere} & $\xi_{RIN, Atmos}$ is the noise component due to the scintillation caused by atmospheric fluctuations. We will quantify this noise in the next section. \\
\hline
$\xi_{RIN, LO}$ & {\it RIN of the LO} &The noise term $\xi_{RIN, LO}$ is due to the power fluctuations of the laser before modulation. This noise component is an intrinsic noise that depends on the parameters of the laser \cite{laud}. \\
\hline
$\xi_{mod}$& {\it Modulation noise} &The noise term $\xi_{mod}$ is due to the voltage fluctuations in the modulation of the coherent state at Alice \cite{filip}. The signal generator translates bit information to a voltage which is amplified to drive the modulator \cite{laud}. The phase of the quadrature is proportional to the applied voltage. Subsequently, the voltage deviation introduced by the signal generator introduces modulation noise. \\

\hline

$\xi_{Background}$ & {\it Background noise} & Part of the channel excess noise for the satellite-to-Earth channel is photon leakage from the background $\xi_{background}$ to the quantum signal and LO  \cite{swang2}. \\ 

%
\hline
$\xi_{RIN, Signal}$ & {\it RIN of the signal} & Excess noise due to power fluctuations of the laser translates to the RIN of the signal $\xi_{RIN, Signal}$. This noise component is proportional to the absolute power variance for a given optical bandwidth of the signal laser \cite{laud}. In comparison to the RIN of the LO due to the laser, the RIN of the signal is conisderably smaller \cite{laud}. \\
 \hline \hline
 $\xi_d$ & Detector excess noise component &  \\
 \hline\hline
$v_{el}$ & {\it Electronic noise} & The term $v_{el}$ is electronic noise that is due to other noise sources of the detector including thermal noise and clock jitter affecting the detector. Shot-noise-limited homodyne measurement requires sufficient LO power to reduce the effect of the electronic noise by increasing the signal-to-noise ratio (SNR) \cite{laud, huangsci}. \\ 
\hline
$\xi_{ADC}$ & {\it Analogue-digital converter (ADC) quantization noise} & $\xi_{ADC}$ is due to digitizing the output voltage, required by all CV-QKD systems \cite{laud, ren, tao}. This noise term can be suppressed by increasing the number of bits.\\


\hline
$\xi_{overlap}$ & {\it Pulse overlap} & The finite response time of the balanced homodyne/heterodyne detector causes an electrical pulse overlap $\xi_{overlap}$ \cite{chi}. Large peak powers for shorter pulses can saturate the diodes, causing a non-linear response. This can be suppressed by ensuring $\frac{1}{\tau_0^2 f_{rep}^2}>>1$, where $\tau_0$ is the pulse width. \\
\hline
$\xi_{LO}$ & {\it LO fluctuations during subtraction} & The incomplete subtraction of the output signals by the homodyne/heterodyne detector introduces the noise component $\xi_{LO}$ \cite{chi, laud}. \\
\hline
$\xi_{Leak}$ & {\it Leakage noise} & Part of the total excess noise is the photon leakage from the LO to the signal when the signal is multiplexed with the LO \cite{huangsci, chi}. This term largely depends on the system design. However, in the polarization-frequency-multiplexing scheme, this is assumed negligible \cite{bqi}. \\
 \hline
 \end{tabular}
 \caption{Noise components of the total excess noise $\xi$ of CS-QKD the satellite-to-Earth channel.}
    \label{table1}
\end{table*}

\begin{table*}[t!]
\centering
 \begin{tabular}{||l l l l l||}
 \hline
 Parameter &  &   & & Ref. \\ [0.5ex]
 \hline\hline
   $\xi_{ch}$ & & Channel excess noise & $0.0186; 0.0126$ & \\
    & $\xi_{ta}$ & Time-of-arrival fluctuations & $0.006$ & This paper \\
              & $\xi_{RIN, Atmos}$ & RIN of LO due to atmosphere & $0.01; 0.003$ & This paper \\
                      & $\xi_{RIN, LO}$ & Relative intensity noise of LO & $0.0018$ &\cite{laud}  \\
             & $\xi_{mod}$ & Modulation noise & $0.0005$ & \cite{filip} \\
          & $\xi_{Background}$ & Background noise & $0.0002$ & \cite{swang2} \\
        & $\xi_{RIN,Signal}$ & Relative intensity noise of signal & $<0.0001$ & \cite{laud}  \\
            \hline
   $\xi_{d}$ && Detector noise & $0.0133$ & \\
 & $v_{el}$ & Electronic noise & $0.013$ & \cite{chi, huangsci} \\
& $\xi_{ADC}$ & Analogue-digital converter noise & $0.0002$ & \cite{ren, tao} \\
& $\xi_{overlap}$ & Detector overlap & $<0.0001$ & \cite{chi} \\
& $\xi_{LO}$ & LO subtraction noise & $<0.0001$ & \cite{laud, chi} \\
 & $\xi_{Leak}$ & LO to signal leakage & $<0.0001$ & \cite{huangsci, chi, bqi} \\
 \hline
 \end{tabular}
 \caption{Excess noise break-down affecting satellite-to-Earth CS-QKD during daylight.}
 \label{table}
\end{table*}

\section{Noise simulations}
In the following section, we use the atmospheric channel to determine the channel excess noise contributions from the intensity fluctuations of the LO and the time-of-arrival fluctuations between the LO and signal. Typical values from the literature of each noise contribution (as well as our calculations) are summarised in Table \ref{table}.
\label{simulate}
\subsection{Relative intensity - Local oscillator}
We follow the procedure presented in \cite{laud} to derive the contribution to the channel excess noise from the variations in the intensity of the LO, in the context of the satellite-to-Earth channel. At Bob, the signal and LO are mixed on a balanced beamsplitter. Photo-detectors at the outputs measure the intensity and are subtracted to obtain the quadrature measurement. The difference in the number operator describes this measurement
\begin{align}
\Delta \hat{n}_{B} = |\alpha_\text{LO}|(\cos(\theta) \hat{X} + \sin(\theta) \hat{P}),
\end{align} where $|\alpha_\text{LO}|$ is the coherent amplitude and $\theta$ is the phase of the quadrature measurement.
It is clear that oscillations in the amplitude $|\alpha_\text{LO}|$ of the LO will introduce an additional excess noise to the protocol. Assuming for simplicity we intend to measure the $\hat{X}$ quadrature ($\theta = 0$). Then the difference in number operator is reduced to
$\Delta \hat{n}_{B} = |\alpha_\text{LO}| \hat{X}$. Since the fluctuations in intensity and the quadratures are independent random variables, then the variance of the difference in number operator is proportional to the variance of both LO and quadrature operator,
\begin{align}
\text{Var}(\Delta \hat{n}_{B}) &= \text{Var}(|\alpha|_{LO} \hat{X}) \\
&= \langle |\alpha_\text{LO}|^2 \rangle \langle \hat{X}^2 \rangle -  \langle |\alpha_\text{LO}| \rangle^2 \langle \hat{X} \rangle^2 \nonumber \\
&=  \langle |\alpha_\text{LO}|^2 \rangle V_A \nonumber \\
&=  (\text{Var}(|\alpha_\text{LO}|) + \langle |\alpha_\text{LO}| \rangle^2) V_A,
\label{eq:dn1}
\end{align} since $\braket{\hat{X}}=0$ and $\text{Var}(|\alpha_{LO}|)=\langle |\alpha_\text{LO}| \rangle^2-\langle |\alpha_\text{LO}|^2 \rangle$ is determined by the intensity fluctuations of the LO and $V_A$ denotes the variance of the quadrature without considering the effects of the variation of intensity in the LO (i.e. Alice's modulation variance).
To derive the variance in the measurement of the $\hat{X}$ quadrature considering all the effects, we rewrite $\text{Var}(\Delta \hat{n}_{B})$ assuming a constant intensity LO with amplitude $\langle |\alpha_\text{LO}| \rangle$, and that the RIN in the quadrature measurement is characterized by the variance $V_{RIN, LO}(\hat{X})$.
\begin{align}
\text{Var}(\Delta \hat{n}_{B}) &= \text{Var}(\langle |\alpha_\text{LO}| \rangle \hat{X}) \nonumber \\
&=  \langle |\alpha_\text{LO}| \rangle^2 V_{total}(\hat{X}) \nonumber \\
&=  \langle |\alpha_\text{LO}| \rangle^2 (V_A + V_{RIN, LO}(\hat{X})).
\label{eq:dn2}
\end{align} where $V_{total}(\hat{X})$ is the total variance including $V_A$ and the variance of the quadrature operator due to the RIN of the LO.
Comparing both (\ref{eq:dn1}) and (\ref{eq:dn2}) we get
\begin{align}
 \langle |\alpha_\text{LO}| \rangle^2  V_{RIN, LO} (\hat{X}) =  \text{Var}(|\alpha_\text{LO}|) V_A.
\end{align}
Finally, the resulting noise component is
\begin{align}
\xi_{RIN} = \frac{\text{Var}(|\alpha_\text{LO}|)}{\braket{|\alpha_\text{LO}|} ^2} V_\text{A}.
\label{eq:noise_atmos}
\end{align}

We identify two main sources in the fluctuations of the intensity of the LO. One is the fluctuations inherent to the laser when it is generated and the second is the fluctuations caused by the atmospheric fading channel. Since these two random effects are independent, then the noise component due to intensity fluctuations of the LO becomes the sum of the two contributions $\xi_{RIN} = \xi_{RIN, LO} + \xi_{RIN, Atmos}$. In \cite{laud} the noise term\footnote{$\xi_{RIN, LO}=\frac{1}{4} RIN_{LO} \times B_{LO} \times V_A$ where $RIN_{LO}~=~1.4\times10^{-7} Hz^{-1}$~is the RIN and $B_{LO}=10$~KHz is the optical bandwidth of the LO \cite{laud}.} due to the fluctuations of the laser is derived and given for a realistic setup with commonly available lasers. The contribution to the total excess noise is $\xi_{RIN, LO} \approx 0.00035 V_\text{A}$. In Table \ref{table}, $\xi_{RIN, LO}$ is shown for a modulation variance of $V_A=5$.

Analysing $\xi_{RIN, Atmos}$, we note that the intensity of the LO is measured over the aperture used by the receiver.
That is, we consider the total power $P$ over the aperture surface $\mathcal{D}$,
 \begin{align}
 P_\mathcal{D} = \iint_{\mathcal{D}} |I(x,y)|^2 dxdy,
 \end{align}
where $I(x,y)$ is the irradiance of the LO expressed in Cartesian coordinates with the aperture centred at $(x,y)=(0,0)$.
In (\ref{eq:noise_atmos}), the term $\frac{Var(|\alpha_{LO}|)}{\braket{|\alpha_{LO}}^2}$ is the normalized intensity variance known as the scintillation index $\sigma_{SI}^2$. We can quantify the scintillation index averaged over the aperture surface $\mathcal{D}$ with diameter $D_R$
\begin{equation}
\sigma_{SI}^2 (D_R) =\frac{\langle P_\mathcal{D}^2 \rangle }{ \langle P_\mathcal{D} \rangle^2} - 1.
\end{equation}
Then the $\xi_{RIN, Atmos}$ noise component can be rewritten as
\begin{equation}
\xi_{RIN, Atmos} = \sigma_{SI}^2(D_R) V_\text{A}.
\end{equation}
The value of $\sigma_{SI}^2(D_R)$ can be obtained for weak atmospheric turbulence by using the atmospheric models presented in the Appendix, and calculating the first and second-order statistical moments of the irradiance \cite{andrews_book1}. The result is

\begin{equation}
\begin{split} 
  &\sigma_{SI}^2(D_R)= 8.70 k^{7/6} (H-h_0)^{5/6}\sec^{11/6}(\zeta) \\
  &\times \text{Re} \int_{h_0}^{H} C_n^2(h) \left[ \left( \frac{k D_R^2}{16 H} + i \frac{h-h_0}{H-h_0} \right)^{5/6} - \left( \frac{kD_R^2}{16H} \right)^{5/6} \right]dh,
  \end{split}
  \label{something}
\end{equation} where $k=\omega_0/c$ is the wavenumber of the laser frequency $\omega_0$ ($c$ is the speed of light), $\zeta$ is the zenith angle, $H$ is the altitude of the satellite at zenith angle $\zeta=0$, $h_0=0$~km is the altitude of the receiver ground station, and $C^2_n(h)$ is the refractive index structure of the atmosphere (see Appendix).
In Fig. \ref{fig:si_d} we present the values of $\sigma^2_{SI}(D_R)$ for different receiver aperture diameters and zenith angles. We see the importance of using a large aperture in the receiver, since not only is it essential for increasing the transmissivity of the channel but also to reduce the fluctuations on intensity which translates to excess noise. The values of $\xi_{RIN, Atmos}$ are calculated for the receiver aperture diameter $D_R=1$~m and $D_R=3$~m at $\zeta=60^\circ$ and altitude $H=500$~km are given in Table \ref{table}. These are $\xi_{RIN, Atmos}=0.01$ and $0.004$, respectively. For the rest of this paper, unless otherwise specified, we use the former value for $D_R=1$~m.
\begin{figure}
\centering
\includegraphics[width=.49\textwidth]{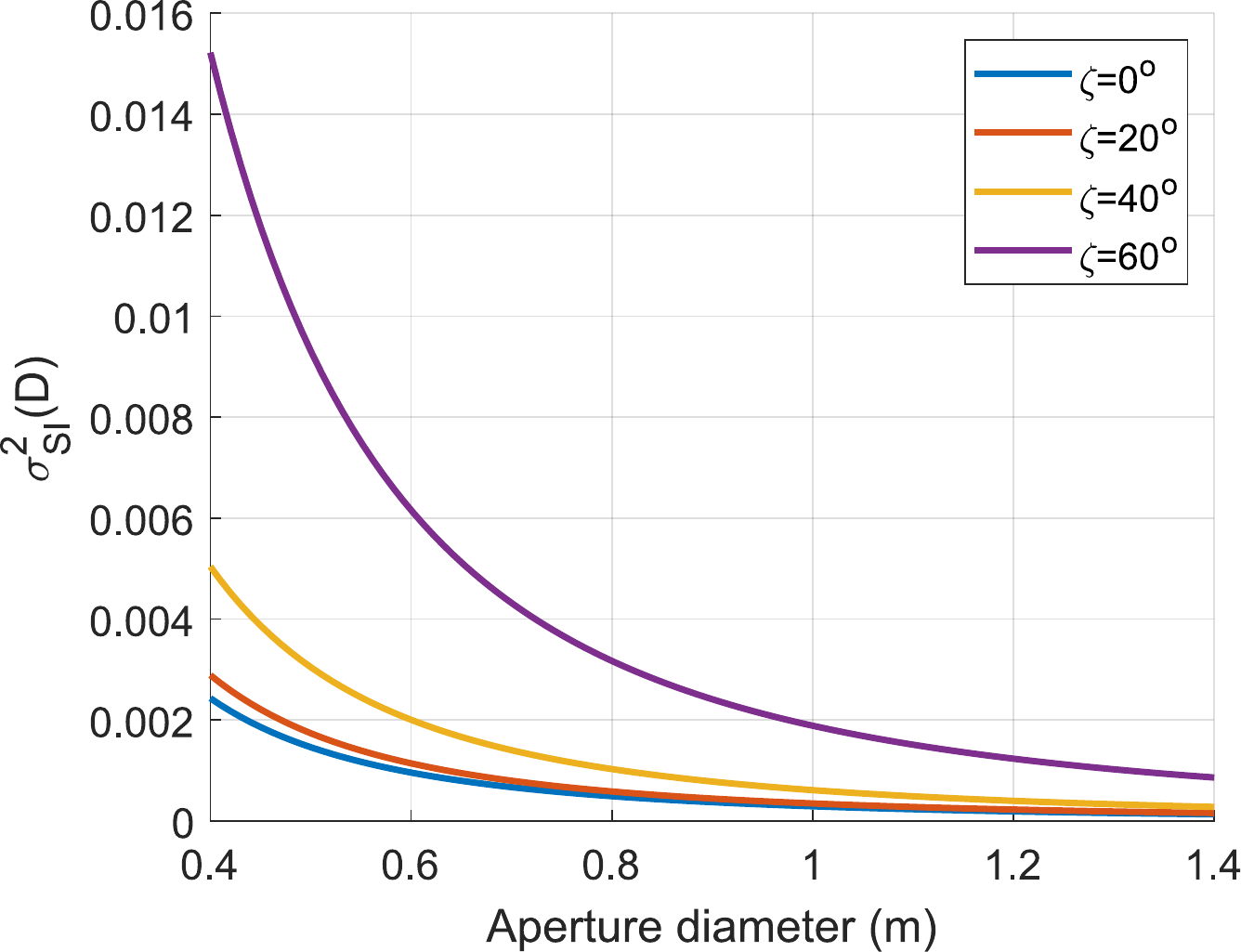}
\caption{Aperture averaged scintillation index for different receiver aperture diameters and zenith angles.}
\label{fig:si_d}
\end{figure}

\subsection{Time-of-arrival fluctuations}
Now we derive the pulse deformations due to transmission through the satellite-to-Earth channel, used in the calculation of the time-of-arrival fluctuations of our Table \ref{table}. In \cite{TemporalBroadening} an analysis based on the mutual coherence function (MCF) is made, where the two-frequency MCF is defined by the average of all ensemble of pulses (see equations (11)-(28) in \cite{TemporalBroadening}).
The MCF represents the product of the free space optical fields and the turbulence factor based on the first and second-order Rytov approximation. Since LEO satellites are positioned at a high enough altitude we can treat the propagation of the wave as in the far-field regime. In this regime the wave as seen by the receiver corresponds to a plane wave.

In the far-field regime corresponding to the Fresnel parameter $\Omega= \omega_0 W_0^2/(2 L c)<<1$ where $W_0$ is the beam-waist, $L$ is the total distance travelled by the pulse, and $c$ is the speed of light. The beam-waist is determined by the transmitter aperture diameter which is a fundamental system parameter, as it dominates over many of the effects occurring during optical propagation. The temporal mean intensity in the far-field regime is obtained by averaging over the bandwidth of the MCF at the same radial position and time. From \cite{TemporalBroadening}, the result is
\begin{align}
\begin{split}
  \langle I(r, L ,t') \rangle &\approx \frac{\tau^2_0 }{\tau_1}\left( \frac{W_0^2}{2Lc} \right)^2 \frac{\tau_0^2 [(1 + \omega_0^2 \tau_0^2) + (W_0 r /Lc)^2]}{\tau_1[\tau_0^2 + (W_0 r /Lc)^2 ]^{5/2}} \\ 
  & \times \exp \left\{ -\frac{\omega_0^2 \tau_0^2(W_0 r /Lc)^2}{2[\tau_0^2 + (W_0 r /Lc)^2 ]}    \right\} \nonumber \\
  &\times \exp \left\{ - \frac{2(t' - L/c - r^2/2Lc)^2}{\tau_1^2} \right\},
\end{split}
  \label{eq:avgI}
\end{align} where $t'$ is the time, $r$ the radial coordinate over the wavefront of the pulse, $\tau_0$ is the initial pulse width, and the broadened pulse width $\tau_1$ is
\begin{align}
  & \tau_1 =  \sqrt{\tau_0^2 + 8\alpha}, \nonumber\\
  & \alpha = \frac{0.391(1+0.171\delta^2 - 0.287\delta^{5/3}) \nu_1 \sec(\zeta)}{c^2},
\end{align}
and where $\nu_1$ is the following integral (which we evaluate numerically),
\begin{equation}
  \nu_1 = \int_{h_0}^{H} C_n^2(h) L_0(h)^{5/3}dh.
\end{equation}
Here $\delta = \frac{L_0}{l_0}$ where $L_0$ and $l_0$ denote the outer and inner scales, respectively. The inner and outer scales are introduced alongside the atmospheric model described in the Appendix.
\begin{figure}
\centering
\includegraphics[width=.49\textwidth]{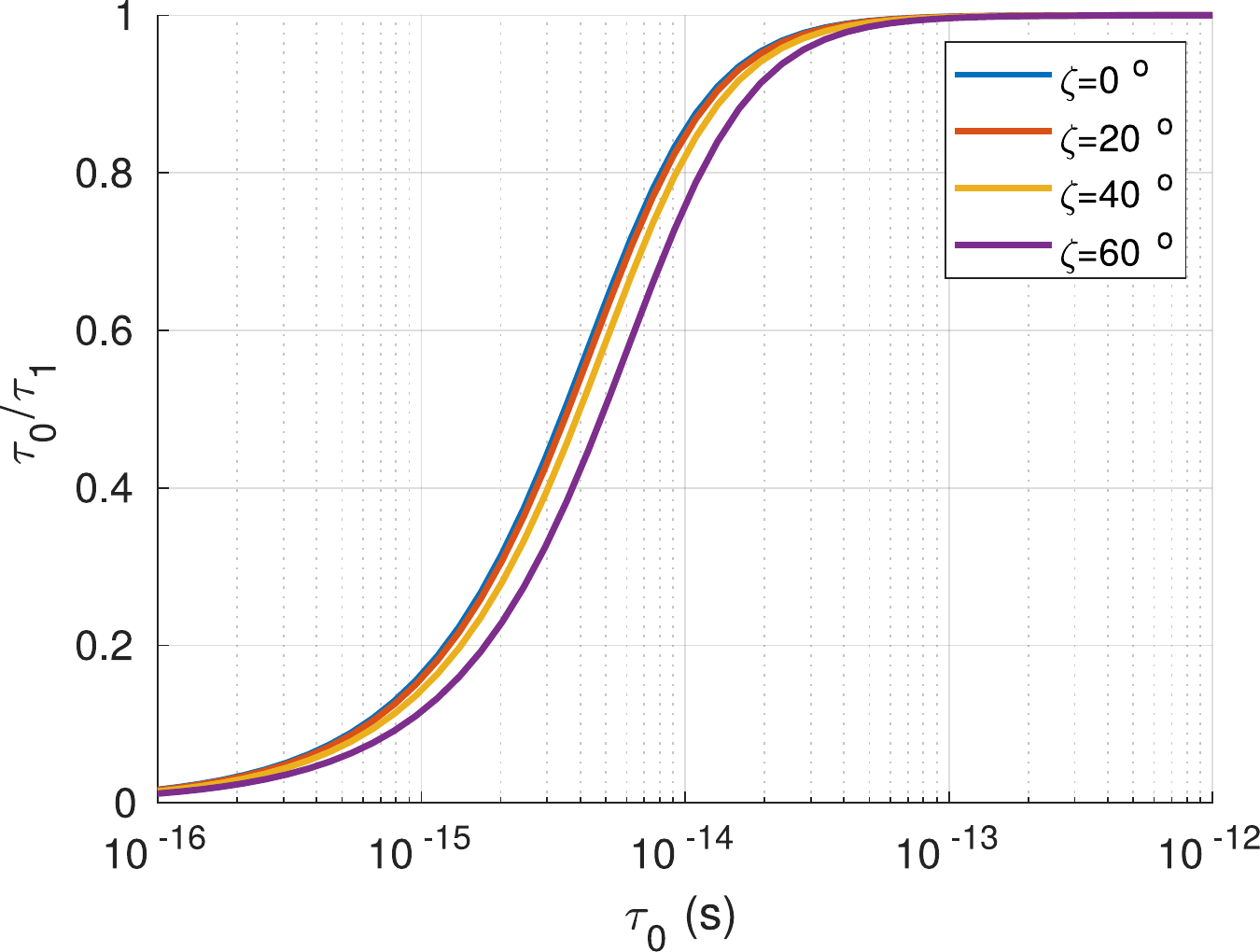}
\caption{Ratio of pulse broadening as a function of initial pulse width and for different zenith angles.}
\label{fig:t1}
\end{figure}
\begin{figure*}[t!]
\centering
\includegraphics[scale=0.4]{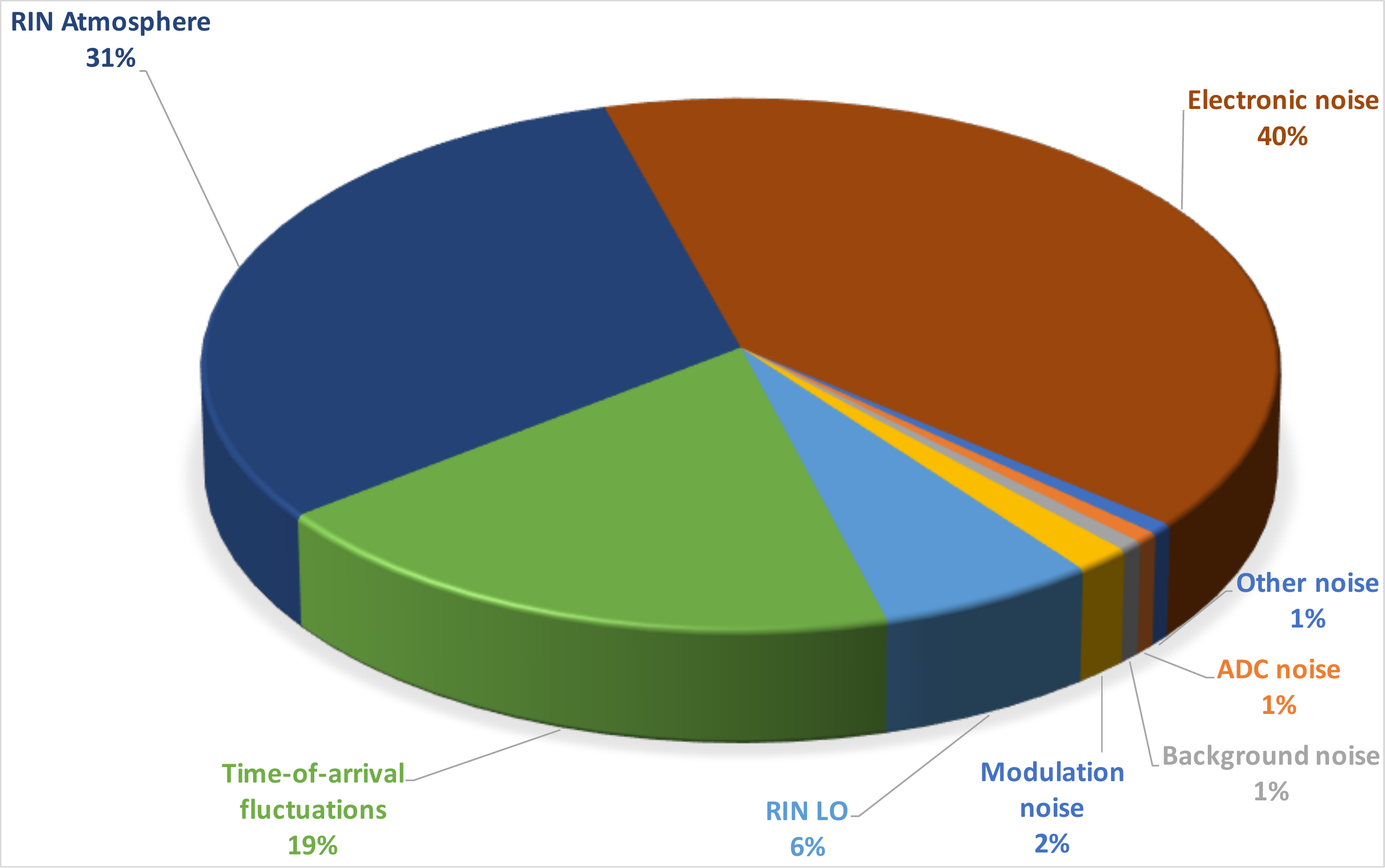}
\caption{Pie chart of noise contributions to $\xi=\xi_{ch}+\frac{\mu \xi_{d}}{\eta_d T}$ for CS-Hom protocol ($\mu=1$) in the satellite-to-Earth channel for daylight operation. We have used values from Table \ref{table} with laser pulse width $\tau_0=130$ $\text{ps}$ and state-of-the-art homodyne detector specifications with $N_{LO}=10^8$. The dominant noise term is the electronic noise at $40\%$, followed by RIN due to the atmosphere at $31\%$, the time-of-arrival fluctuations at $19\%$ and the RIN of the LO at $6\%$.}
\label{piechart}
\end{figure*}
Any optical pulse sent thought through the atmosphere, will incur deformations due to variations of the refractive index of atmospheric turbulence. To fully analyse the pulse deformations we consider two effects; the broadening of the pulse, and the changes on the time-of-arrival of the pulse. Both of these effects are characterised by the value of $\tau_1$.
First, the broadening of the pulse will cause an attenuation of the average light intensity of the received signal by a factor of $\tau_0 / \tau_1$ \cite{swang}.
In Fig. \ref{fig:t1} we show the ratio $\tau_0 / \tau_1$ as a function of $\tau_0$ for the satellite-to-Earth channel with the parameters specified above.
We see that the pulse broadening becomes considerable only for pulse widths $\tau_0 < 0.1 \text{ ps}$. However, for a pulse width $\tau_0=130$~ps commonly used in CV-QKD systems consistent with a $100$~MHz repetition rate \cite{huangsci}, the pulse broadening effect is negligible.

 The second and more relevant effect is the variation of the time-of-arrival $t_a$ at the detector between the signal and the LO pulses. As shown in \cite{ta_fluctuations}, the statistical properties of the time-of-arrival are obtained from the temporal statistical moments of the complex envelope describing the pulse. The analysis shows that the time-of-arrival mean and variance equal $\langle t_a \rangle = L/c$ and $\sigma_{ta}^2= \frac{\tau_1^2}{4}$, respectively.
In satellite-to-Earth CS-QKD, the fluctuations of the time-of-arrival caused by the atmospheric channel contribute to the total excess noise as an additional amount \cite{swang}
\begin{equation}
\xi_{ta}=2 V_A \omega_0^2 (1-\rho_{ta}) \sigma_{ta}^2,
\label{toaf}
\end{equation} where $\rho_{ta}$ is the timing correlation coefficient between LO and the signal. We set the correlation coefficient between signal and LO to $\rho_{ta}=1-10^{-13}$ as in \cite{swang}.
The noise component due to time-of-arrival fluctuations (\ref{toaf}) increases with the pulse width $\tau_1$. In Table \ref{table}, we determine the value for this noise component to be  $\xi_{ta}=0.006$ for a pulse width of $\tau_0=130$~ps and central frequency $\omega_0=2 \pi \times 200$~THz (corresponding to a wavelength of $1550$~nm).


\begin{figure}[!htb]
    \centering
    \begin{minipage}{.49\textwidth}
        \centering
\includegraphics[scale=0.85]{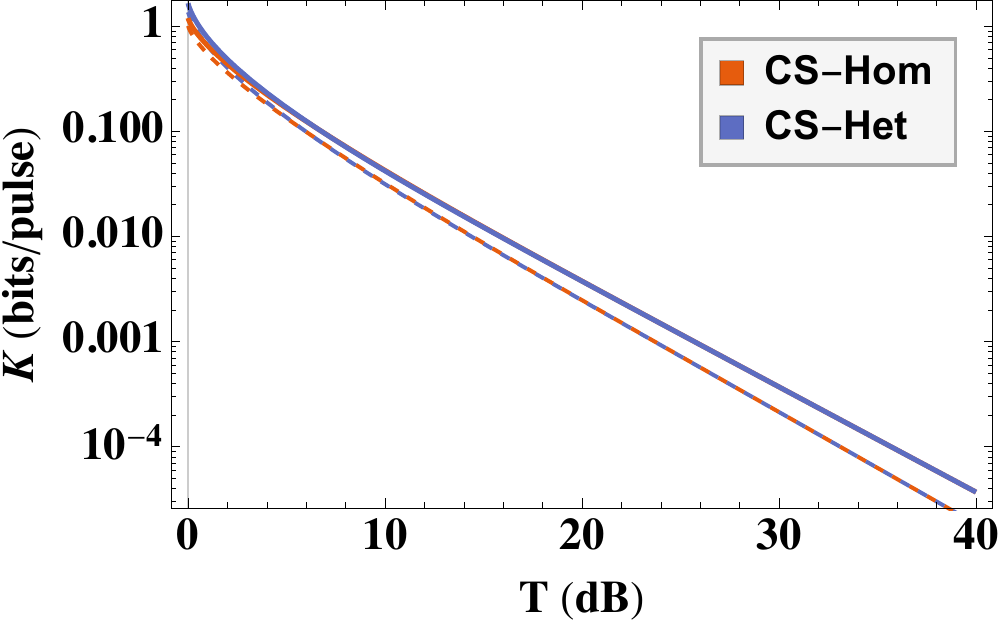}
\caption{Secret key rate in the asymptotic limit plotted against transmissivity for CS-Hom (red) vs. CS-Het (blue) for different channel excess noises $\xi_{ch}=0$ (solid) and our chosen satellite-to-Earth channel (dashed).}
\label{skrloss}
\end{minipage}%
\hfill
    \begin{minipage}{0.49\textwidth}
\centering
\includegraphics[scale=0.87]{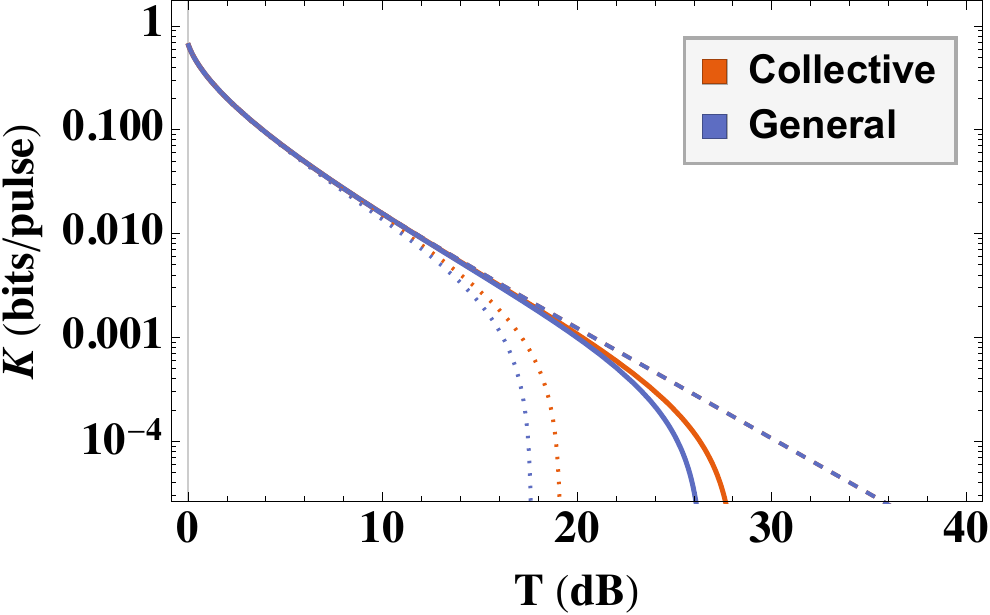}
\caption{Secret key rate in the finite limit for the CS-Het protocol for collective and general attacks plotted against transmissivity for different $n=10^{10}$ (dotted line), $n=10^{12}$ (solid line) and the asymptotic limit (dashed line).}
\label{skrlossfinite}
   \end{minipage}
\end{figure}


The LO generated by Alice must have sufficient intensity at Bob such that he can perform a heterodyne/homodyne measurement. Typically, $10^8$ photons per pulse are required at Bob's side. With a $100$ $\text{MHz}$ pulse repetition rate and pulse width of $130$~ps, the required LO average power for a $20$~dB channel loss (a typcial loss considered here) at Alice's side is $130$~mW (at $1550$~$\text{nm}$). We have chosen the repetition rate to coincide with current detector technology \cite{huangsci}. There are readily available coherent laser sources with these specifications. We assume the ground station has a state-of-the-art homodyne detector as in \cite{chi}. 

In Fig. \ref{piechart} we compare the noise components contribution for daylight operation. As $T$ gets smaller, the term $\frac{\mu \xi_{d}}{\eta_d T}$ in (\ref{ex}) would dominate, but for the purpose of this figure, we compare each noise component on equal footing by setting $T=1$ and $\eta_d=1$.  In this figure, we adopt the following values for the noise components. The electronic noise $v_{el}=0.013$ is the dominant noise component. The next dominant noise component is the relative intensity noise due to the atmosphere. For the purpose of the noise analysis, we assume the realistic scenario, where the satellite zenith angle is $\zeta=60^{\circ}$ (which is sufficient for a $2-3$ minutes flyby \cite{liao}), a transmitter aperture diameter of $D_T=0.3$~m and a receiver aperture diameter of $D_R=1$ m with an altitude $H=500$~km, implying a relative intensity noise of $\xi_{RIN,atmos}=0.01$.

The next dominant noise component is the time-of-arrival fluctuations. We set the correlation coefficient between signal and LO to $\rho_{ta}=1-10^{-13}$ as in \cite{swang}. Fig. \ref{fig:t1} suggests that for pulse widths $\tau_0 > 0.1$ ps, the pulse broadening due to weak turbulence is negligible and $\tau_1\approx \tau_0$. Thus, the excess noise due to time-of-arrival fluctuations is $\xi_{ta}=0.006$. As expected, this noise component doesn't depend on the receiver aperture diameter.

Furthermore, $\xi_{RIN, LO}$ is $6\%$ of the total excess noise. ADC noise is negligible ($1\%$) as part of the detector noise component. All other noises $\xi_{RIN, Signal}$, $\xi_{LO}$ and $\xi_{Leak}$ are less than $1\%$ of the total excess noise with the exception of the modulation noise which is $2\%$ \cite{laud}. Therefore, the predicted channel excess noise is $\xi_{ch}=0.0186$ and the detector excess noise $\xi_{d}=0.0133$.

\begin{table}[b!]
\begin{tabular}{| l | l | l | l | l | l | l | l | l | l |}
\hline
$D_R$ &$H$ & $\zeta$ & $\eta_d$ & $\beta$ & $V_A$ & $\tau_0$ & $\lambda$ & $\xi_{ch}$&$\xi_d$ \\
\hline
$1$~m&$500$~km &$60^{o}$ &$0.95$ & $0.95$ & $5$ & $130$~ps & $1550$~nm &$0.0186$ &$0.0133$ \\
\hline
\end{tabular}
\caption{System parameters.}
\label{system}
\end{table}

\section{Secret key rates of satellite-to-Earth CS-QKD}
\label{impact}
In Fig.~\ref{skrloss} we calculate the secret key rate in the asymptotic limit against the transmissivity $T$ for both CS-Hom and CS-Het protocols using the parameters in Table \ref{system}. Unless otherwise specified, for the rest of the paper, we use the parameters in Table  \ref{system}. The transmissivity on the x-axis is converted to units of dB using the formula $-10 \times \log_{10}{T}$. We compare the channel excess noises of $\xi_{ch}=0$ and that of our specified satellite-to-Earth channel. For the chosen parameter values, corresponding to a realistic experimental  scenario, the difference in the secret key rate between CS-Hom and CS-Het is only evident at low loss in which the CS-Het protocol performs marginally better. The modulation variance $V_A$ is an adjustable parameter that can be optimized to maximize the secret key rate with the total excess noise and transmissivity \cite{laud}. However, in atmospheric channels  the transmissivity can fluctuate (although here we assume not) and the optimal value of $V_A$ at one time may not be the optimal value at another time. Unless the goal is to maximize the secret key rate each run, additional resources to optimize $V_A$ are not necessary. In our system, we choose a value of $V_A=5$ which is in the range of optimality for the values of transmissivity in the satellite-to-Earth channel.

In Fig.~\ref{skrlossfinite} we calculate the secret key rate in the finite case against the transmissivity $T$. As seen in Fig.~\ref{skrlossfinite} for the CS-Het protocol, the loss that can be tolerated (compared to the asymptotic limit) is less in the finite size regime for $n=10^{10}$ and $n=10^{12}$. In this same figure, we compare the key rates for collective and general attacks using the failure probability $\epsilon=10^{-9}$ and $\epsilon=10^{-55}$ ($\epsilon'=10^{-9}$), respectively. It is evident that the general attack does not tolerate as much loss. Note it is straightforward to convert all key rate into units of bits/s by multiplying by the laser pulse repetition frequency.\footnote{In real-world  
crytopgraphy it is the key rate in bits/s that is ultimately of interest. However, conversion from bits/pulse to bit/s must consider the time to decode the finite, but very large, block lengths (multiplying by the pulse repetition rate is only formally valid  when the reconciliation time is zero). Optimisation of the bit/s key rate can be non-trivial in some circumstances \cite{xai}.}

In  Table \ref{tablekey} the individual impact of each noise term on the secret key rate under collective Gaussian attacks is shown\footnote{Note in our calculations we have assumed that the transmissivity remains constant over the block length.  Receiving a block length of 10$^{12}$ from a 100MHz source would require minutes of transmission time. For block lengths larger than 10$^{10}$ it is likely that the impact of satellite motion on transmissivity would need to be considered within the calculations unless a corresponding higher-rate source was utilised.} in the finite size regime with $n=10^{12}$.

\begin{table*}[t!]
\centering
 \begin{tabular}{||l l l l l l||}
 \hline
 Parameter & Description &  Value & $\frac{K}{K_{0}}$ @ $10$ dB & $\frac{K}{K_{0}}$ @ $20$ dB & $\frac{K}{K_{0}}$ @ $30$ dB\\ [0.5ex]
 \hline\hline
   $\xi_{ch}$ & Channel excess noise & $0.0186$ & $0.75$ & $0.63$ & $0.00$\\
     $\xi_{ta}$ & Time-of-arrival fluctuations & $0.006$ & $0.91$ & $0.87$ & $0.00$ \\
      $\xi_{RIN, Atmos}$ & Relative intensity noise due to atmosphere & $0.01$ & $0.85$ & $0.79$ & $0.00$ \\
           $\xi_{Background}$ & Background noise & $0.0002$ & $1.00$ & $1.00$ & $0.93$ \\
   $\eta_d$ & Detector efficiency & $0.95$ & $0.95$ &$0.95$ & $0.95$ \\
  $v_{el}$ & Electronic noise & $0.013$ & $0.99$ & $0.99$ & $0.81$ \\[1ex]
 \hline
 \end{tabular}
 \caption{Individual impact of each noise term on secret key rate under collective Gaussian attacks relative to ideal case of no noise with $K$ in the finite size regime (i.e. the ratio $\frac{K}{K_0}=1$ for no impact and $\frac{K}{K_0}=0$ for maximum impact). Other parameter values: $n=10^{12}$, $\epsilon=10^{-9}$, modulation variance $V_A=5$ and reconciliation efficiency $\beta=0.95$.}
 \label{tablekey}
\end{table*}
The background noise during daylight is approximately $1\%$ of the total excess noise for $1550$ nm light ($0.0002$) \cite{swang2}. In DV-QKD, background noise from daylight can be mitigated using small spatial filtering coupled with advanced adaptive optics \cite{grun}. It was noted in \cite{bed} that optical systems used in CV-QKD could also use small spectral bandwidth. Consequently, the background noise can be filtered, allowing for daylight operation. Daylight CV technology was demonstrated in \cite{gunt}, but there are no corresponding studies for CV-QKD.

We find that the background noise component during daylight in CS-QKD in the satellite-to-Earth channel has a negligible impact on the secret key rate, in agreement with \cite{swang2}. For night-time operation, the secret key rate would be unchanged with $\xi_{Background}<10^{-7}$. Thus, the secret key rate in Fig. \ref{skrlossfinite} is a lower bound, and night-time operation would improve the key rates, but is not essential.

We consider how increasing the size of the receiver aperture improves the secret key rates under general attacks. For a larger receiver aperture of $D_R=3$~m, we obtain a smaller channel excess noise of $\xi_{ch}=0.0126$ and detector excess noise of $\xi_{d}=0.0133$. At the transmissivity $T=15$~dB, a secret key rate of $2.6 \times 10^{-3}$~bits/pulse is attained under general attacks for this larger receiver aperture. $T=15$~dB is readily achievable in the satellite-to-Earth channel using this larger receiver aperture.
For example, for the transmitter aperture of $D_T=0.3$~m at transceiver separation $1200$~km the beam width at the ground is $12$~m (and a diffraction-limited divergence of $10$~$\mu$rad), giving a diffraction loss of $22$~dB for receiver aperture $D_R=1$~m (see discussion below for Micius).
Extrapolating from these parameters, the receiver aperture diameter of $D_R=2.3$~m gives a diffraction loss of $15$~dB. For this receiver aperture of $D_R=2.3$~m, the channel excess noise is $\xi_{ch}=0.015$ and a secret key rate of $2.4 \times 10^{-3}$~bits/pulse is attained under general attacks. 

We consider how the secret key rate changes with the system parameters in Table \ref{system}. For example, doubling the pulse period to $\tau_0=260$~ps, will give a channel excess noise of $\xi_{ch}=0.033$. However, the secret key rate is barely changed (i.e. $1.0 \times 10^{-3}$~bits/pulse). We also find that from $V_A=2$ to $V_A=8$, the result doesn't significantly change. Hence, this main result is not sensitive to the parameters we have chosen for the figures.

As we grew close to the completion of our study, two other works appeared; one \cite{neda2020} that can be made applicable to satellite-to-Earth channels and one \cite{deq} directly applicable to these channels. Different from our study, these latter works had a focus on the large-scale fading effects on the QKD key rates, and different excess noise contributions. They, therefore, apply to different system models than the diffraction-only loss models we explored here. However,
  collectively the works of \cite{neda2020,deq} and ours illustrate clearly that CV-QKD in the satellite-to-Earth channel is feasible.

Although our focus has been to demonstrate the viability of the CS-QKD protocols in satellite-to-Earth channel,  it is perhaps interesting to compare our results directly with Micius.
In our calculations of the secret key rates, we have assumed that transmissivity is constant.\footnote{The statistical nature of the turbulence only manifests itself in our calculations through the excess noise contributions. In particular, this is through the impact of scintillation on $\xi_{RIN, Atmos}$ and on the random refractive index perturbations on $\xi_{ta}$.  Under the assumption of diffraction-only losses, the constant transmissivity  $T$ of the channel is determined via the system parameters, $D_T$, $D_R$ and $L$. Of course, these parameters must be set so that the diffraction-only loss assumption is a reasonable one.}
That is,  transmissivity fluctuations due to atmospheric turbulence and beam-wandering are assumed small in comparison to diffraction losses \cite{kau}.
As just mentioned above, in the Micius satellite QKD experiment, the satellite-to-Earth channel loss due to diffraction was $22$~dB at a maximum distance of $1200$~$\text{km}$ with a transmitter aperture of $D_T=0.3~$m, a far-field divergence of $10$~$\mu\text{rad}$ and receiver aperture of $D_R=1$~m \cite{liao}. Additional loss occurred via atmospheric absorption and turbulence   ($3$ to $8$~dB), and  loss due to pointing error  ($<3$~dB). This gave an average total loss of $30\pm 3$~dB.
In calculations of the  Micius QKD key rate, the failure probability was set to  $\epsilon=10^{-9}$ which gave a key rate of $1.38 \times 10^{-5}$ bits/pulse. In comparison, for our CS-Het protocol at the same security setting and same losses we find zero key rate under general attacks. It is only for losses less than 25~dB that we find non-zero rates under general attacks. We do caution however, that direct comparisons of these two very different technologies (DV vs. CV) are limited in value because the security assumptions are different. We also point out the Micius security analysis disregards information leakage due to side-channel imperfections \cite{liao}. Other assumptions underpinning the Micius key rate are discussed in \cite{lih}.
However, to be fair, we have also overlooked some assumptions that underpin our CV-QKD analysis. The most important of these are in regard to the LO - an issue we discuss next.

\section{LO Hacking}
\label{qhacking}
\subsection{Attacks}
An underlying assumption implicit in key rates we have determined for the CS-QKD protocols is that the transmission of the LO through the satellite-to-Earth channel is secure and not interfered with. However, clearly this assumption is suspect as an eavesdropper can in principle obtain access to the LO, modify the channel, and subsequently obtain information on the quantum key. Here we briefly discuss this important caveat. We mention some specific attacks and some countermeasures.

{\it Equal-amplitude attack}. In this attack, Eve intercepts both the LO and quantum signal and performs homodyne detections on both \cite{has}. Eve produces two squeezed states with the same intensity level. Bob cannot discover the attack because the expected noise is less than the shot noise level. When Bob compares the difference in his measurement with Alice's prepared state to the threshold from theoretical security proofs, he mistakenly confirms the channel is secure \cite{gross1, lod}. Eve has therefore compromised the channel. However, this security loophole can be solved if Bob can monitor the fluctuations in the intensity of the LO \cite{ma}.

{\it Wavelength attack}. In this attack, Eve modifies the intensity transmission of Bob's beamsplitter by changing the wavelength of the light from Alice. This wavelength attack deems the final keys insecure. However, it was shown in \cite{hchrist}, that adding a simple wavelength filter randomly before measuring the intensity of the LO can prevent this attack.

{\it Calibration attack}. The device calibration routine of the LO is susceptible to an intercept-resend attack. In this attack, Eve manipulates the LO pulses to change the clock pulses used for detection  \cite{paul}. More specifically, Eve changes the shape of the LO pulse which consequently induces a delay in the clock trigger. The homodyne detection will measure a lower intensity. Therefore, there will be a decrease in the response slope between the variance of the homodyne detection and the LO power. As a result, the shot noise will be overestimated and the excess noise underestimated \cite{jzhuang}. Countermeasures to this attack involves monitoring the shot noise in real time.

%

\subsection{Local local oscillators}
Motivated by these security loopholes, there has recently been a focus on using \textit{true local oscillators} or \textit{local local oscillators} (LLO). 
In such systems, two lasers are used, one at Alice for generating the quantum signal and another at Bob for the LLO. Reference pulses are sent with the signals which are used to calibrate the phase offset in the LLO \cite{ren, koashi, huang, marie}. However, the phase offset derived from the reference pulses is limited by quantum uncertainty, and the two lasers necessarily introduce some phase drift. In comparison to optical fibre, no phase compensation method exists for the LLO in FSO channels where additional phase noise due to atmospheric turbulence is expected \cite{qi}, but these are not insurmountable issues. Despite their intuitive appeal over LO schemes, attack strategies against LLO schemes do exist. In one such attack recently proposed \cite{wzhao}, Eve intercepts the reference pulse and estimates the phase drift using Bayesian algorithms. Consequently, the excess noise is biased to mask an intercept-resend attack and a beam-splitter attack.


In general, any pilot symbol scheme (e.g. reference pulses) is always subject to manipulation by an adversary, and therefore susceptible to attacks outside the scope of current security proofs \cite{scar}. Formal security analyses that account for the optimal general attack on both quantum signals and pilot symbols are needed to close this security gap. An alternative approach would be the use of a system that avoids the use of any reference signals at all for the phase referencing. Only by this means do the assumptions underpinning formal security proofs for the CV-QKD protocols we have discussed here remain intact. Recent work along these lines is given in \cite{kleis} where publicly announced raw key values are invoked to determine the phase offsets to be applied to the LLO. However, currently even this type of phase determinations for the LLO are not formally proven secure - an issue that hopefully will be addressed in forthcoming works. 

\section{Discussion}
\label{discuss}
 In our work, we have deliberately focused on the feasibility of the simplest-to-deploy CS-QKD protocols over the satellite-to-Earth channel for which composable security is known. The reason for choosing the simplest-to-deploy CS-QKD protocols was two-fold. First, early CV-based satellites will certainly deploy one of these protocols, if not both. Second, if the easiest-to-deploy protocols were shown to lead to close-to-zero key rates then more sophisticated protocols (with likely lower key rates) would likely be not interesting.

Our work largely focused on the on CS-Het protocol for which composable security in the finite key limit (i.e, a deployable scenario) is known. However, it is worth mentioning that other CV-QKD protocols exist for which composable security in the finite limit is also known, such as two-mode squeezed state CV-QKD protocol \cite{furrer}. These other protocols in the finite-key setting are generally either more difficult to deploy or have lower expected key rates, relative to CS-Het.

An approach to preventing side-channel attacks against detectors is available via CV measurement device-independent (CV MDI-QKD) protocols. These types of CV-QKD protocols allow for untrusted devices that act as relays. Interestingly, formal proofs of security in the finite regime are known for CV MDI-QKD \cite{lupo, mdi, zhengyu}. An improvement to this protocol,  the coherent-state-based ``Twin-Field" (TF)-QKD protocol (an extension of \cite{lucamar}), has been recently proposed, e.g. \cite{xma, yin} and proof-of-principle experiments carried out in optical fibre \cite{zhong}.  Information-theoretic security in the finite regime for coherent state TF-QKD has also been proposed \cite{yin2}. However, we do note that in scenarios where the untrusted relay is a satellite being served in the uplink, both MDI-QKD and TF-QKD will suffer losses substantially larger than the downlink losses. Phase referencing between the uplink signals sent from Alice and Bob (and authentication of each pulse so referenced - at the cost of key consumption), and phase compensation of the signals traversing the different FSO uplink channels, must also be included. All these issues will reduce the final key rate.


\section{Conclusion}
\label{conclusion}
In this work, we investigated the feasibility of the simplest-to-deploy CV-QKD protocols, the CS-QKD protocols, over the satellite-to-Earth channel. We reviewed
the security of these two protocols, as well as the components of the total excess noise most relevant to the satellite-to-Earth channel. Through new calculations we then determined that the most important sources of noise in our protocols over the satellite-to-Earth channel was the relative intensity noise due to the atmosphere, and the time-of-arrival fluctuations between the signal and LO. We next focused on the CS-QKD protocol with heterodyne detection in the deployable setting of a finite key regime.
 Given reasonable transceiver aperture sizes, and assuming diffraction as the only source of loss, we found that  reasonable QKD key rates under general attacks can be anticipated. We, therefore, concluded that CV-QKD with information-theoretic security in the satellite-to-Earth channel is entirely feasible.
In closing, we noted our information-theoretic security assumes that the LO is secured from manipulation by an adversary.
Future useful work in this area could include formal security proofs of finite-limit CS-QKD with homodyne detection, and formal proofs of security under LO (or reference pulse) attacks.

\section*{Acknowledgments}
This research was a collaboration between the Commonwealth of Australia (represented by the Defence Science and Technology Group) and the University of New South Wales through a Defence Science Partnerships agreement.

\newpage

\newpage
\pagebreak
\section*{Appendix: Atmospheric models}

Turbulence in the Earth's atmosphere is caused by random variation in temperature and pressure. These variations alter the air's refractive index both spatially and temporally, distorting any optical waves propagating through the atmosphere. Distortions of the pulses transmitted such as beam deformation and beam wandering are one of the main sources of photon loss in the channel.
To model the effects of the turbulent atmosphere on the optical signals we use Kolmogorov's theory of turbulent flow \cite{Kolmogorov}. The strength of the turbulence is characterized by the refractive index structure parameter, $C_n^2$. Kolmogorov's theory is based upon the insight that turbulence is induced by eddies in the atmosphere. The scale of the turbulence can be characterised by an inner-scale $l_0$, and an outer-scale $L_0$ \cite{andrews_book1}. Eddies with a size lower than the inner-scale cannot exist since the turbulence at such scales is dissipated into the atmosphere as heat. The outer-scale denotes the upper-bound to the eddy size and is the scale at which energy is injected into the turbulence.

To model the atmospheres vertical profile of the refractive index structure parameter, we use the widely used Hufnagel-Valley model \cite{HVmodel},
\begin{align}
C_n^2(h) &= 0.00594(v/27)^2 (10^{-5}h)^{10} \exp(-h/1000) \\ \nonumber
          &+ 2.7\times10^{-16} \exp(-h/1500) + A \exp(-h/100),
\end{align}
with $h$ the altitude in meters, $v=21$ the rms wind-speed (m/s), and $A=1.7 \times 10^{-14}$ the nominal value of $C_n^2(0)$ at the ground.
Additionally, measurements made of the scintillation suggest the outer scale $L_0$ changes with the altitude according to the empirical Coulman-Vernin profile \cite{outer_scale}
\begin{align}
L_0(h) = \frac{4}{1 + \left(\frac{h-8500}{2500}\right)^2}.
\end{align} Finally, the inner scale is assumed to directly proportional to the outer scale $l_0=\delta L_0$, with $\delta=0.005$ \cite{TemporalBroadening}.

To describe the fluctuations of the refractive index we use a spectral density function \cite{book_phase_screen}
\begin{align}
\Phi_{\phi}(\kappa) = 0.49 r_0^{-5/3} \frac{\exp(-\kappa^2/\kappa^2_m)}{{(\kappa^2 + \kappa_0^2)}^{11/6}},
\label{pdf}
\end{align}
with $\kappa$ the radial spatial frequency on a plane orthogonal to the propagation direction, $\kappa_m = 5.92/l_0$, $\kappa_0= 2\pi/L_0$, and $r_0$ the Fried parameter for a vertical propagation length \cite{andrews_book1}
\begin{align}
r_0 = \left( 0.423 k^2 \sec(\zeta)\int_{h_0}^{H} C_n^2(h) dh \right)^{-3/5},
\end{align}
where $h_0$ and $H$ correspond to the lower and upper altitudes of the propagation path.

%
%



\begin{thebibliography}{99}

\bibitem{letter} A. Boaron et al., {\it Secure Quantum Key Distribution over 421 km of Optical Fiber}, Phys. Rev. Lett. {\bf 121}, 190502 (2018).
\bibitem{liao} S.-K. Liao et al., {\it Satellite-to-ground Quantum Key Distribution}, Nature (London) {\bf 549}, 43 (2017).

\bibitem{bed} R. Bedington, J. M. Arrazola, A. Ling, {\it Progress in Satellite Quantum Key Distribution}, NPJ Quantum Inf., vol. {\bf 3}, no. 30, pp. 1-13 (2017).

\bibitem{ralph} T. C. Ralph, {\it Continuous Variable Quantum Cryptography}, Phys. Rev. A, {\bf 61}, 010303 (1999).

\bibitem{hill} M. Hillery, {\it Quantum Cryptography with Squeezed States}, Phys. Rev. A, {\bf 61}, 022309 (2000).

\bibitem{reid} M. D. Reid, {\it Quantum Cryptography with a Predetermined Key, using Continuous-variable Einstein-Podolsky-Rosen Correlations}, Phys. Rev. A, {\bf 62}, 062308 (2000).

\bibitem{cerf} N. J. Cerf, M. Levy, G. van Assche, {\it Quantum Distribution of Gaussian Keys with Squeezed States}, Phys. Rev. A {\bf 63}, 052311 (2001).

\bibitem{gmod} F. Grosshans, G. van Assche, J. Wenger, R. Brouri, N. J. Cerf, P. Grangier, {\it Quantum Key Distribution Using Gaussian-Modulated Coherent States}, Nature {\bf 421} 238 (2003).

\bibitem{gross} F. Grosshans, P. Grangier, {\it Continuous Variable Quantum Cryptography Using Coherent States}, Phys. Rev. Lett. {\bf 88}, 057902 (2002).


\bibitem{gross2} F. Grosshans, P. Grangier, {\it Reverse Reconciliation Protocols for Quantum Cryptography with Continuous Variables}, Proceedings of the 6th International Conference on Quantum Communications, Measurement, and Computing (2002).

\bibitem{weedbrook} C. Weedbrook, A.M. Lance, W.P. Bowen, T. Symul, T. C. Ralph, P.K. Lam, {\it Quantum Cryptography without Switching}, Phys. Rev. Lett. {\bf 93}, 170504 (2004).

\bibitem{xwang} X. Wang, S. Guo, P. Wang, W. Liu, and Y. Li, {\it Realistic Rate-distance Limit of Continuous-variable Quantum Key Distribution}, Opt. Express {\bf 27}, 13372-13386 (2019).

\bibitem{jou} P. Jouguet, S. Kunz-Jacques, A. Leverrier, P. Grangier, E. Diamanti, {\it Experimental Demonstration of Long-distance Continuous-variable Quantum Key Distribution}, Nat. Photonics {\bf 7}, 378-381 (2013).

\bibitem{ber} G. P. Berman, A. A. Chumak, {\it Photon Distribution Function for Long-distance Propagation of Partially Coherent Beams Through the Turbulent Atmosphere}, Phys. Rev. A {\bf 74}, 013805 (2015).

\bibitem{usenko} V. C. Usenko, B. Heim, C. Peutinger, C. Wittmann, C. Marquardt, G. Leuchs, R. Filip, {\it Entanglement of Gaussian States and the Applicability to Quantum Key Distribution Over Fading Channels}, New J. Phys. {\bf 14}, 093048 (2012).

\bibitem{heim}  B. Heim, C. Peuntinger, N. Killoran, I. Khan, C. Wittmann, C. Marquardt, G. Leuchs, {\it Atmospheric Continuous-variable Quantum Communication}, New J. Phys. {\bf 16}, 113018 (2014).

\bibitem{qu} Z. Qu, I. B. Djordjevic, {\it Four-dimensionally Multiplexed Eight-state Continuous-variable Quantum Key Distribution over Turbulent Channels}, IEEE Photon. J., vol. {\bf 9}, no. 6, (2017).

\bibitem{gong} Y.-H. Gong, K.-X. Yang, H.-L. Yong, J.-Y. Guan, G.-L. Shentu, C. Liu, F.-Z. Li, Y. Cao, J. Yin, S.-K. Liao, J.-G. Ren, Q. Zhang, C.-Z. Peng, and J.-W. Pan, {\it Free-space Quantum Key Distribution in Urban Daylight with the SPGD Algorithm Control of a Deformable Mirror}, Opt. Express {\bf 26}, 18897 (2018).


\bibitem{papan} P. Papanastasiou, C. Weedbrook, S. Pirandola, {\it Continuous-variable Quantum Key Distribution in Uniform Fast-fading Channels}, Phys. Rev. A {\bf 97}, 032311 (2018).


\bibitem{rupert} L. Ruppert, C. Peuntinger, B. Heim, K. Gnthner, V. C. Usenko, D. Elser, G. Leuchs, R. Filip and C. Marquardt, {\it Fading Channel Estimation for Free-space Continuous-variable Secure Quantum Communication}, New J. Phys. {\bf 21}, 123036 (2019).

\bibitem{derk} I. Derkach, V. C. Usenko, R. Filip, {\it Squeezing-enhanced Quantum Key Distribution Over Atmospheric Channels}, New J. Phys. {\bf 22}, 053006 (2020).

\bibitem{polarize} S.Y. Shen, M.W. Dai, X.T. Zheng, Q.Y. Sun, B. Zhu, G.C. Guo, Z.F. Han, {\it Free-Space Continuous-variable Quantum Key Distribution of Unidimensional Gaussian Modulation using Polarized Coherent-states in Urban Environment}, Phys. Rev. A {\bf 100}, 012325 (2019).

\bibitem{neda1} N. Hosseinidehaj, Z. Babar, R. Malaney, S. X. Ng and L. Hanzo, {\it Satellite-Based Continuous-variable Quantum Communications: State-of-the-Art and a Predictive Outlook}, IEEE Communications Surveys \& Tutorials, vol. {\bf 21}, no. 1, pp. 881-919 (2019).

\bibitem{laud} F. Laudenbach, C. Pacher, C-H. Fred Fung, A. Poppe, M. Peev, B. Schrenk, M. Hentschel, P. Walther, H. Hubel {\it Continuous-variable Quantum Key Distribution with Gaussian Modulation - The Theory of Practical Implementations}, Adv. Quantum Technol. Article 1800011 (2018).

\bibitem{virtual}  F. Grosshans, N. J. Cerf, J. Wenger, R. Tualle-Brouri, and P. Grangier, {\it Virtual Entanglement and Reconciliation Protocols for Quantum Cryptography with Continuous Variables}, Quantum Inf. Comput. {\bf 3}, 535 (2003).


\bibitem{diam} E. Diamanti, A. Leverrier, {\it Distributing Secret Keys with Quantum Continuous Variables: Principle, Security and Implementations}, Entropy {\bf 17}, 6072 (2015).

\bibitem{renner} R. Renner, J. I. Cirac, {\it de Finetti Representation Theorem for Infinite-Dimensional Quantum Systems and Applications to Quantum Cryptography}, Phys. Rev. Lett. {\bf 102}, 110504 (2009).


\bibitem{lev2} A. Leverrier, {\it Security of Continuous-variable Quantum Key Distribution via a Gaussian de Finetti Reduction}, Phys. Rev. Lett. {\bf 118}, 200501 (2017).

\bibitem{foss}  S. Fossier, E. Diamanti, T. Debuisschert, R. Tualle Brouri, and P. Grangier, {\it Improvement of Continuous-variable Quantum Key Distribution Systems by Using Optical Preamplifiers}, J. Phys. B {\bf 42}, 114014 (2009).


\bibitem{dev} I. Devetak, A. Winter, {\it Distillation of Secret Key and Entanglement from Quantum States}, Proc. R. Soc. A. {\bf 461}: 207-235 (2005).

\bibitem{scar} V. Scarani, H. Bechmann-Pasquinucci, N. J. Cerf, N. Du\v{s}ek, Norbert L\"{u}tkenhaus, M. Peev, {\it The Security of Practical Quantum Key Distribution}, Rev. Mod. Phys. {\bf 81}, 1301 (2009).

\bibitem{gross1} F. Grosshans and N. J. Cerf, {\it Continuous-variable Quantum Cryptography is Secure Against Non-Gaussian Attacks}, Phys. Rev. Lett. {\bf 92}, 047905 (2004).

\bibitem{garcia}  R. Garcia-Patron, N. J. Cerf, {\it Unconditional Optimality of Gaussian Attacks against Continuous Variable Quantum Key Distribution}, Phys. Rev. Lett. {\bf 97}, 190503 (2006).

\bibitem{takaya} T. Matsuura, K. Maeda, T. Sasaki, M. Koashi, {\it Finite-size Security of Continuous-variable Quantum Key Distribution with Digital Signal Processing}, arXiv:2006.04661v1 (2020).

\bibitem{lev} A. Leverrier, F. Grosshans, P. Grangier, {\it Finite-size Analysis of Continuous-variable Quantum Key Distribution}, Phys. Rev. A {\bf 81}, 062343 (2010).


\bibitem{imperfect} P. Jouguet, S. Kunz-Jacques, E. Diamanti, A. Leverrier, {\it Analysis of Imperfections in Practical Continuous-variable Quantum Key Distribution}, Phys. Rev. A {\bf 86}, 032309 (2012).


\bibitem{levc} A. Leverrier, {\it Composable Security Proof for Continuous-variable Quantum Key Distribution with Coherent States}, Phys. Rev. Lett. {\bf 114}, 070501 (2015).


\bibitem{lupo} C. Lupo, C. Ottaviani, P. Papanastasiou, S. Pirandola, {\it Continuous-variable Measurement-Device-Independent Quantum Key Distribution: Composable Security Against Coherent Attacks}, Phys. Rev. A {\bf 97}, 052327 (2018).


\bibitem{neda2020} N. Hosseinidehaj, N. Walk, T. C. Ralph, {\it Composable Finite-size Effects in Free-space CV-QKD Systems}, arXiv:2002.03476 (2020).

\bibitem{yu} D. Yu. Vasylyev, A. A. Semenov, W. Vogel, {\it Atmospheric Quantum Channels with Weak and Strong Turbulence}, Phys. Rev. Lett. {\bf 117}, 090501 (2016).

\bibitem{Eduardo}  E. E. Villase\~{n}or, R. Malaney, K. A. Mudge, K. J Grant, {\it Atmospheric Effects on Satellite-to-ground Quantum Key Distribution using Coherent States},  arXiv:2005.10465 (2020).

\bibitem{filip} V. C. Usenko, R. Filip, {\it Feasibility of Continuous-variable Quantum Key Distribution with Noisy Coherent States}, Phys. Rev. A {\bf 81}, 022318 (2010).

\bibitem{swang2} S. Wang, P. Huang, T. Wang, G. Zeng, {\it Feasibility of All-Day Quantum Communication with Coherent Detection}, Phys. Rev. Applied {\bf 12}, 024041 (2019).


\bibitem{huangsci} D. Huang, P. Huang, D. Lin, and G. Zeng, {\it Long-distance Continuous-variable Quantum Key Distribution by Controlling
Excess Noise}, Sci. Rep. {\bf 6}, 19201 (2016).


\bibitem{ren} S. Ren, R. Kumar, A. Wonfor, X. Tang, R. Penty and I. White, {\it Noise and Security Analysis of Trusted Phase Noise Continuous Variable Quantum Key Distribution using a Local Local Oscillator}, 2019 IEEE 20th International Workshop on Signal Processing Advances in Wireless Communications (SPAWC), Cannes, France, pp. 1-5 (2019).

\bibitem{tao} T. Wang, P. Huang, Y. Zhou, W. Liu, G. Zeng, {\it Pilot-Multiplexed Continuous-variable Quantum Key Distribution with a Real Local Oscillator}, Phys. Rev. A 97, 012310 (2018).


\bibitem{chi} Y.-M. Chi, B. Qi, W. Zhu, L. Qian, H.-K. Lo, S.-H. Youn, A. I. Lvovsky, L. Tian, {\it A Balanced Homodyne Detector for High-Rate Gaussian-Modulated Coherent-State Quantum Key Distribution}, New J. Phys. {\bf 13}, 013003 (2011).

\bibitem{bqi} B. Qi, L.-L. Huang, L. Qian, H.-K. Lo, {\it Experimental Study on the Gaussian-modulated Coherent-state Quantum Key Distribution over Standard Telecommunication Fibers}, Phys. Rev. A {\bf 76}, 052323 (2007).












\bibitem{andrews_book1} L. C. Andrews, R. L. Phillips, {\it Laser Beam Propagation through Random Media}, SPIE Press, Second edition (2005).

\bibitem{TemporalBroadening} D. E. T. T. S. Kelly, L. C. Andrews, {\it Temporal Broadening and Scintillations of Ultrashort Optical Pulses}, Waves in Random Media, {\bf 9} (1999).

\bibitem{swang} S. Wang, P. Huang, T. Wang, and G.-H. Zeng, {\it Atmospheric Effects on Continuous-variable Quantum Key Distribution}, New J. Phys. {\bf 20}, 083037 (2018).

\bibitem{ta_fluctuations} C. Y. Young, L. C. Andrews, A. Ishimaru, {\it Time-of-arrival Fluctuations of a Space-time Gaussian Pulse in Weak Optical Turbulence: An Analytic Solution}, Appl. Opt. {\bf 37}, 7655-7660 (1998).

\bibitem{xai} X. Ai, R. Malaney and S. X. Ng, {\it A Reconciliation Strategy for Real-time Satellite-based QKD}, in IEEE Communications Letters, vol. {\bf 24}, no. 5, pp. 1062-1066 (2020).

\bibitem{grun} M. T. Gruneisen, B. A. Sickmiller, M. B. Flanagan, J. P. Black, K. E. Stoltenberg, and A. W. Duchane, {\it Adaptive spatial Filtering of Daytime Sky Noise in a Satellite Quantum Key Distribution Downlink Receiver}, Opt. Eng. {\bf 55}, 026104 (2016).

\bibitem{gunt} K. G\"{u}nthner, I. Khan, D. Elser, B. Stiller, \"{O}. Bayraktar, C. M\"{u}ller, K. Saucke, D. Tr\"{o}ndle, F. Heine, S. Seel, P. Greulich, H. Zech, B. G\"{u}tlich, S. Philipp-May, C. Marquardt, and G. Leuchs, {\it Quantum-limited Measurements of Optical Signals From a Geostationary Satellite}, Optica  {\bf 4}, 611-616 (2017).

\bibitem{deq} D. Dequal, L. T. Vidarte, V. R. Rodriquez, G. Vallone, P. Villoresi, A. Leverrier, E. Diamanti, {\it Feasibility of Satellite-to-ground Continuous-variable Quantum Key Distribution}, 	arXiv:2002.02002 (2020).


\bibitem{kau} H. Kaushal, G. Kaddoum, {\it Optical Communication in Space: Challenges and Mitigation Techniques}, IEEE Commun. Surveys Tuts., vol. {\bf 19}, no. 1, pp. 57-96 (2017).

\bibitem{lih} Li, H., Yin, Z., Wang, S. et al. {\it Randomness Determines Practical Security of BB84 Quantum Key Distribution}, Sci Rep {\bf 5}, 16200 (2015).

\bibitem{has} H. H\"{a}seler, T. Moroder, and N. L\"{u}tkenhaus, {\it Testing Quantum Devices: Practical Entanglement Verification in Bipartite Optical Systems}, Phys. Rev. A, {\bf 77}, 032303 (2008).

\bibitem{lod} J. Lodewyck, P. Grangier, {\it Tight Bound on the Coherent-state Quantum Key Distribution with Heterodyne Detection}, Phys. Rev. A. {\bf 76}, 022332 (2007).

\bibitem{ma} X.-C. Ma, S.-H. Sun, M.-S. Jiang, L.-M. Liang, {\it Local Oscillator Fluctuation Opens a Loophole for Eve in Practical Continuous-variable Quantum-key-distribution Systems}, Phys. Rev. A {\bf 88}, 022339 (2013).


\bibitem{hchrist} J.-Z. Huang, C. Weedbrook, Z.-Q. Yin, S. Wang, H.-W. Li, W. Chen, G.-C. Guo, Z.-F. Han, {\it Quantum Hacking of a Continuous-variable Quantum-key-distribution System using a Wavelength Attack}, Phys. Rev. A {\bf 87}, 062329 (2013).

\bibitem{paul} P. Jouguet, S. Kunz-Jacques, E. Diamanti, {\it Preventing Calibration Attacks on the Local Oscillator in Continuous-variable Quantum Key Distribution}, Phys. Rev. A {\bf 87}, 062313 (2013).

\bibitem{jzhuang} J.-Z. Huang, S. Kunz-Jacques, P. Jouguet, C. Weedbrook, Z.-Q. Yin, S. Wang, W. Chen, G.-C. Guo, Z.-F. Han, {\it Quantum Hacking on Quantum Key Distribution using Homodyne Detection}, Phys. Rev. A {\bf 89}, 032304 (2014).













\bibitem{koashi} M. Koashi, {\it Unconditional Security of Coherent-State Quantum Key Distribution with a Strong Phase-Reference Pulse}, Phys. Rev. Lett. {\bf 93}, 120501 (2004).


\bibitem{huang} D. Huang, P. Huang, D. Lin, C. Wang, G. Zeng, {\it High-speed Continuous-variable Quantum Key Distribution Without Sending a Local Oscillator}, Opt. Lett. {\bf 40}, 3695 (2015).




\bibitem{marie} A. Marie, R. All\'{e}aume, {\it Self-coherent Phase Reference Sharing for Continuous-variable Quantum Key Distribution}, Phys. Rev. A {\bf 95}, 012316 (2017).

\bibitem{qi} B. Qi, P. Lougovski, R. Pooser, W. Grice, M. Bobrek, {\it Generating the Local Oscillator Locally in Continuous-variable Quantum Key Distribution Based on Coherent Detection}, Phys. Rev. X {\bf 5}, 041009 (2015).

\bibitem{wzhao} W. Zhao, R. Shi, D. Huang, {\it Practical Security Analysis of Reference Pulses for Continuous-variable Quantum Key Distribution}, Scientific Reports {\bf 9}, 18155 (2019). 


\bibitem{kleis} S. Kleis, M. Rueckmann, C. G. Schaeffer, {\it Continuous-variable Quantum Key Distribution with a Real Local Oscillator and without Auxiliary Signals}, 	arXiv:1908.03625 (2019).

\bibitem{furrer} F. Furrer, T. Franz, M. Berta, A. Leverrier, V. B. Scholz, M. Tomamichel, R. F. Werner, {\it Continuous Variable Quantum Key Distribution: Finite-Key Analysis of Composable Security against Coherent Attacks}, Phys. Rev. Lett. 109, 100502 (2012); Erratum Phys. Rev. Lett. 112, 019902 (2014).



\bibitem{mdi} S. Pirandola, C. Ottaviani, G. Spedalieri, C. Weedbrook, S. L. Braunstein, S. Lloyd, T. Gehring, Christian S. Jacobsen,  U. L. Andersen, {\it High-rate Quantum Cryptography in Untrusted
Networks}, Nature Photon. {\bf 9}, 397 (2015).

\bibitem{zhengyu} Z. Li, Y.-C. Zhang, F. Xu, X. Peng, H. Guo, {\it Continuous-variable Measurement-Device-Independent Quantum Key Distribution}, Phys. Rev. A {\bf 89}, 052301 (2014).

\bibitem{lucamar} M. Lucamarini, Z. Yuan, J. F. Dynes, A. J. Shields, Overcoming the Rate Distance Limit of Quantum Key Distribution Without Quantum Repeaters, Nature {\bf 557}, 400-403 (2018).

\bibitem{xma} X. Ma, P. Zeng, H. Zhou, Phase-matching Quantum Key Distribution, Phys. Rev. X. {\bf 8}, 031043 (2018).


\bibitem{yin} H.-L. Yin, Z.-B Chen, {\it Coherent-state-based Twin-field Quantum Key Distribution}, Sci. Rep. {\bf 9}, 14918 (2019).

%

\bibitem{zhong} X. Zhong, J. Hu, M. Curty, L. Qian, H.-K. Lo, {\it Proof-of-Principle Experimental Demonstration of Twin-Field Type Quantum Key Distribution}, Phys. Rev. Lett. {\bf 123}, 100506 (2019).

\bibitem{yin2}  H.-L. Yin, Z.-B Chen, {\it Finite-key Analysis for Twin-Field Quantum Key Distribution with Composable Security}, Scientific Reports {\bf 9}, 17113 (2019).



%
%

\bibitem{Kolmogorov} A. N. Kolmogorov et al., {\it The Local Structure of Turbulence in Incompressible Viscous Fluid for Very Large Reynolds Numbers}. Proceedings of the Royal Society of London {\bf 434} (1991).

\bibitem{HVmodel} R. E. Hufnagel, N. R. Stanley, {\it Modulation Transfer Function Associated with Image Transmission Through Turbulent Media}, J. Opt. Soc. Am. {\bf 54} (1964).

\bibitem{outer_scale} C. E. Coulman et al., {\it Outer Scale of Turbulence Appropriate to Modeling Refractive-index Structure Profiles}, Appl. Opt. {\bf 27} (1988).

\bibitem{book_phase_screen} J. D. Schmidt, {\it Numerical Simulation of Optical Wave Propagation with Examples in MATLAB}, SPIE Press (2010).

\end{thebibliography}
\end{document}